\documentclass[12pt,a4paper]{article}

\usepackage{titlesec}
\titleformat{\section}{\normalfont\large\bfseries}{\thesection}{1em}{}
\titleformat{\subsection}{\normalfont\normalsize\bfseries}{\thesubsection}{1em}{}
\usepackage{amsmath,amssymb,amsthm}
\usepackage{hyperref}
\usepackage{microtype}
\usepackage{enumitem}
\usepackage{booktabs}
\usepackage{xcolor}
\usepackage{natbib}
\usepackage{array}
\usepackage{rotating}
\usepackage{tabularx}
\usepackage{float}

\usepackage[letterpaper,
            margin=1in]{geometry}

\hypersetup{colorlinks=true, linkcolor=blue!70!black, citecolor=green!60!black}

\usepackage{multirow}
\usepackage{xcolor}

\hypersetup{colorlinks=true,
            linkcolor=blue!70!black}

\theoremstyle{definition} 
\newtheorem{theorem}{Theorem}
\newtheorem{proposition}{Proposition}
\newtheorem{corollary}{Corollary}
\newtheorem{lemma}{Lemma}
\newtheorem{remark}{Remark}
\theoremstyle{definition}

\newcommand{\R}{\mathbb{R}}
\newcommand{\E}{\mathbb{E}}
\newcommand{\Var}{\mathrm{Var}}

\newcommand{\Rbar}{\bar{R}}
\newcommand{\vM}{\mathrm{vM}}
\newcommand{\vMF}{\mathrm{vMF}}
\newcommand{\IG}{\mathrm{IG}}
\newcommand{\lc}{\ell_c}
\newcommand{\lPois}{\ell_{\mathrm{Pois}}}
\newcommand{\lGauss}{\ell_{\mathrm{Gauss}}}
\newcommand{\lIG}{\ell_{\mathrm{IG}}}
\newcommand{\lvM}{\ell_{\vM}}

\newcommand{\kap}{\kappa}
\newcommand{\lam}{\lambda}

\newcommand{\Sd}{\mathcal{S}^{d-1}}

\newcommand{\Sone}{\mathcal{S}^{1}}
\newcommand{\II}{\mathcal{I}}

\newcommand{\wG}{\widetilde{\Gamma}}
\newcommand{\Id}{\mathrm{I}}
\newcommand{\IA}{\mathrm{A}}

\newcommand{\rmin}{r_{\rm min}}
\newcommand{\rmax}{r_{\rm max}}

\newcommand{\Cab}{C_{\rmin,\rmax}}

\newcommand{\bu}{\mathbf{u}}
\newcommand{\bs}{\mathbf{s}}

\newcommand{\beps}{\boldsymbol{\epsilon}}

\newcommand{\K}{\mathcal{K}}

\title{\textbf{Bayesian Inference  for Multivariate Stable Distributions
  via LePage Representation with
  von Mises -- Fisher Spectral Measure}}
\author{Noonhil Kizhakkemadam Unnikrishnan}
\date{}

\begin{document}

\maketitle
\begin{abstract}
We develop Bayesian inference for multivariate $\alpha$-stable distributions with
von Mises--Fisher spectral measure, motivated by the hitting
distribution of Brownian motion on the unit sphere.
Inference proceeds via a LePage-type series representation with latent
radial, spectral, and Poisson components; truncation bias is absorbed
through a Gaussian error term, yielding a likelihood that is smooth in
$\alpha\in(0,2)$ with no special treatment at the Cauchy boundary
$\alpha=1$.
The stability index $\alpha$, spectral concentration $\kappa$, mean
direction $\eta$, and spectral mass $\Gamma_\infty$ are estimated
jointly by Markov chain Monte Carlo.
 At the Gaussian boundary $\alpha\to 2$, the limiting covariance matrix
equals twice the second moment matrix of the spectral measure, with
explicit parametric form determined by $\eta$ and $\kappa$; in
particular, the anisotropy of the limiting Gaussian distribution is
controlled solely by the spectral concentration $\kappa$.
Simulation across $\alpha\in[0.25, 1.99]$ confirms reliable recovery
of all parameters.
\end{abstract}
\medskip

\noindent\textbf{Keywords:}
Brownian motion; Cauchy boundary;
Gaussian boundary;  L\'{e}vy-Khintchine
integral; Markov chain Monte Carlo; spectral measure induced
covariance matrix.

\section{Introduction}

Stable distributions are increasingly being used in both theoretical and empirical investigations in applied literature. Starting with a well cited article by \citet{Mandelbrot1963}, there has been empirical support for thick tail laws in general and stable laws in particular. \citet{McFarlandPettitSung1982}  used symmetric stable laws to study returns from foreign exchange market; as a continuation of this paper, \citet{So1987} observed that return distribution follow negatively skewed stable laws. \citet{Buckle1995} in his analysis of return of a share price observed negatively skewed stable laws. \citet{Tsionas1999}
used sub-Gaussian univariate stable laws in a stable regression setup. \citet{CarrWu2003} used negatively skewed stable laws in analyzing persistence of
volatility smirk across maturity spectrum in the context of S\&P 500 Index
options; they observed that a skewed univariate stable law does a better
job of fitting option prices as compared to stochastic volatility
models. \citet{MitnikRachev2000} has extensively used thick tails distributions. Thus, skewed univariate stable law has been receiving the attention of researchers in empirical finance.

There has been increased attention towards multivariate stable distributions in the last two decades. Here again, majority of the applications that we are aware of are from applied finance, mainly with the objective of modeling financial returns from a portfolio. Commonly studied examples are from foreign exchange markets and stock markets, such as, joint movement of a basket
of currencies or a set of stock indices.  Also, there are occasions where distribution of the basket as a whole is necessary; computation of Value at Risk (VaR) of a portfolio is one such example. Most computations of VaR are currently based on multivariate Gaussian laws, which is known to underestimate losses. Limitations of multivariate Gaussian laws led researchers to practical alternatives, such as historical simulation methods \citep{Jorion2007, BradelyTaqqu2003}, and copula functions \citep{BradelyTaqqu2003, CherubiniLucianoVecchiato2004}. 

An alternative strand of literature approaches this problem using sub-Gaussian models; see, for instance, \citet{QiouRavishanker1998}. However, as elucidated by \citet{SamorodnitskyTaqqu1994}, sub-Gaussian models are 'not very far away from Gaussian models' in the sense of Proposition 2.5.7 \citep{SamorodnitskyTaqqu1994}. Moreover, not all symmetric stable laws are sub-Gaussian.

If individual return distribution of each item in a basket is likely to have thick tail, as most research with univariate laws suggests, it is more natural to use a multivariate law with this property. Due to generalised central limit theorem, multivariate stable laws are ideal candidates, but are also notorious for their mathematical intractability. Multivariate stable laws are implicitly described through its characteristic function which involves a measure on the unit sphere in ${\mathcal R}^d$, known as spectral measure. Basic properties and tail dependence structures of multivariate stable laws are expressed in terms of the spectral measure.

Despite its importance in understanding multivariate stable laws, statistical estimation of spectral measure continues to be a major challenge. Attempts in estimating spectral measure so far have been based on a discrete approximation result by \cite{ByczkowskiNolanRajput1993}  and results on univariate projections by \citet{AbdulHamidNolan1998}. Research by Nolan and his collaborators has given us reasonable understanding of bivariate stable laws; see, for instance, \citet{NolanPanorska1997} for bivariate stable densities in foreign exchange rates. \citet{PivatoSeco2003} approached the problem of estimation of spectral measure
through spherical harmonic analysis.  More recently, \citet{Tsionas2016} developed Bayesian inference using one-dimensional projections of \citet{MatsuiTakemura2009}.

In the present paper, we approach
the problem using a LePage-type
representation \citep{LePage1980,
LePage1981} with the von Mises -- Fisher 
distribution as the spectral measure.
This approach is motivated by five
factors.

\medskip
\noindent\textbf{von Mises through
  Brownian motion:}
\citet{GordonHudson1977} showed that
the von Mises distribution arises as
the hitting distribution of a
two-dimensional Brownian motion with
drift; see also \citet{GordonHudson1979}.
The pole-seeking Brownian motion of
\citet{DGKendall1974} gives a vivid
physical example through bird
navigation.  To our knowledge, the
present work is the first to apply
the Gordon--Hudson characterisation
to motivate the spectral measure of
a multivariate stable distribution.
\citet{Gatto2013} extended the
Gordon--Hudson result to higher
dimensions.

\medskip
\noindent\textbf{Regular variation:}
For $d$-dimensional stable laws,
the tail probability and directional
behaviour in any cone are jointly
determined by the normalised spectral
measure \citep{Resnick1987,
BasrakDavisMikosch2002}.  The von
Mises -- Fisher distribution provides an
explicit parametric form for this
measure, with the mean direction
$\eta$ identifying the dominant tail
direction and the concentration
$\kap$ measuring tail anisotropy.

\medskip
\noindent\textbf{LePage-type
  representation:}
\citet{DamienLaudSmith1995} developed
an approximation theorem for
simulating one-dimensional infinitely
divisible distributions.  We extend
their approach to the multivariate
setting, identifying a latent
structure within stable
laws --- radial, spectral, and
Poisson components --- that yields
a representation theorem in the
spirit of \citet{LePage1981} and
forms the basis of the Gibbs sampler
(Section~\ref{sec:construction_likelihood}).

\medskip
\noindent\textbf{Truncation error
  through Gaussian:}
The LePage series is truncated at
$m$ terms for numerical implementation,
introducing an approximation error.
We model this error through a
Gaussian component, which
yields an explicit likelihood function
for all parameters jointly --- a
step that makes full Bayesian
inference tractable for $d$-dimensional 
stable distributions.

\medskip
\noindent\textbf{Radial component
  as latent variable:}
The classical approach integrates
out the radial component analytically,
yielding a characteristic function
that requires special treatment at
the Cauchy boundary $\alpha=1$.
In the present paper, the radial
component is retained as an explicit
latent variable in the Gibbs sampler,
following \citet{Kuelbs1973}.  This
has two consequences; the likelihood
is smooth in $\alpha\in(0,2)$ with
no special case at $\alpha=1$; and
the radial draws carry direct
information about $\alpha$ through
a sufficient statistic, independently
of the spectral parameters.  This
is a fundamental departure from
existing approaches to multivariate
stable inference.

Use of von Mises -- Fisher distribution as a model for normalized spectral measure allows us to explore tools of directional statistics  in an altogether different area of $d$-dimensional stable laws. In particular, terminologies of directional statistics elicit natural and lucid interpretation of the mean direction of the spectral mass. Through von Mises -- Fisher distribution, asymmetry of the spectral measure is  controlled by its concentration parameter. 
We believe that this approach eliminates some of the limitations of discrete approximations and also seems to enhance the horizon of understanding of multivariate stable laws. 

A perhaps surprising finding is that
the Gaussian limit of the bivariate
stable distribution with von Mises
spectral measure has an explicit
covariance matrix
$\Sigma_\Gamma=\Gamma_\infty(\Id_2+R_2(\kap)
\IA(\eta))$, where $R_2(\kap)=
I_2(\kap)/I_0(\kap)$ is the
second-order mean resultant length and $\Gamma_\infty$ is the total spectral mass.
This connects the tail anisotropy
of the stable distribution directly
to the Gaussian covariance structure
at the boundary. For general result in $d$-dimension, see Section~\ref{sec:gaussian_boundary}.

The paper is organised as follows. Section~2 motivates the von Mises spectral measure through its connection to the hitting distribution of Brownian motion on the unit circle.
Section~\ref{sec:lepage_truncation} develops the LePage-type representation and complete-data likelihood. Identifiability of the stability index $\alpha$ is established in Section~\ref{sec:alpha_identification}.
Section~\ref{sec:initialisation_diagnostics}
 describes the Markov chain
initialisation, with full details in the online Appendix. Section~6 presents the simulation design and results. Section~7 concludes. 

\section{Directional structure of multivariate stable distributions}

\subsection{Gordon--Hudson--Gatto Theorem}
\label{sec:GH_theorem}

\citet{GordonHudson1977} showed that the von Mises distribution
arises as the hitting distribution of a drifted Brownian motion
on $\Sone$.  \citet{Gatto2013} extended this result to
general dimension $d \geq 2$; the von Mises--Fisher distribution
$\mathrm{vMF}(\mu,\kappa)$ on $\Sd$ is the hitting distribution
of a $d$-dimensional Brownian motion with constant drift
$\nu = \kappa\mu$ on the unit sphere $\Sd$.

\begin{theorem}[Gordon--Hudson--Gatto]\label{thm:GH}
  Let $\{B_t\}_{t \geq 0}$ be a $d$-dimensional standard
  Brownian motion starting at the origin, with constant drift vector
  $\nu=\kappa\mu \in \mathcal{R}^d$, where $\mu \in \Sd$ and
  $\kappa \geq 0$.  That is,
  $B_t = W_t + \kappa \mu\,t$,
  with $W_t$ a standard Brownian motion in $\mathcal{R}^d$.
  Let $\tau = \inf\{t > 0 : \|B_t\| = 1\}$ be the first
  hitting time of $\Sd$, and let
  $S = B_\tau / \|B_\tau\| \in \Sd$
  be the hitting direction.  Then $S$ has the von Mises -- Fisher
  distribution:
  \[
    S \sim \mathrm{vMF}(\mu,\kappa),
  \]
  with density
  \begin{equation}\label{eq:vmf_density}
    f_{\mathrm{vMF}}(s \mid \mu,\kappa)
    = C_d(\kappa)\,\exp(\kappa\,\mu^\top s),
    \qquad s \in \Sd,
  \end{equation}
  where $C_d(\kappa) = \kappa^{d/2-1}/\{(2\pi)^{d/2}
  I_{d/2-1}(\kappa)\}$ is the normalising constant and
  $I_p(\cdot)$ is the modified Bessel function of the first kind
  of order $p$.  
  \end{theorem}
At $d=2$, \eqref{eq:vmf_density} reduces to
  the von Mises density \eqref{eq:vM_density}:
  \begin{equation}\label{eq:vM_density}
    f_{\vM}(\theta \mid \eta,\kap)
    = \frac{e^{\kap\cos(\theta-\eta)}}{2\pi I_0(\kap)},
    \qquad \theta\in[0,2\pi),
  \end{equation}
  where $I_0(\kap)$ is the modified Bessel function of
  the first kind of order zero.

This result motivates our choice of spectral measure for
$X \sim \mathrm{Stable}(\alpha, \Gamma)$ in $\mathcal{R}^d$.
Let  $\wG = \Gamma / \Gamma_\infty$ be the normalised version of the spectral measure $\Gamma$, where $\Gamma_\infty = \Gamma(S^{d-1})$. By setting $\widetilde{\Gamma}  \sim \mathrm{vMF}(\mu,\kappa)$,
the spectral measure acquires a natural Brownian motion
interpretation --- it is the distribution of exit directions
of an independent, drifted Brownian motion from the unit sphere.
The mean direction $\mu$ identifies the dominant tail direction
of the stable distribution and the concentration $\kappa$
measures the degree of tail anisotropy.
To the best of our knowledge, this connection between the vMF
spectral measure and the drifted Brownian motion hitting
distribution has not previously appeared in the stable
distribution literature.

Table~\ref{tab:BM_Spectral_Interpretation} provides the
parameter-level linkage for general $d$.

\begin{table}[H]
\centering
\caption{Linkage between the vMF spectral measure parameters
  and the first hitting direction of a $d$-dimensional drifted
  Brownian motion on $\Sd$
  \citep{GordonHudson1977, Gatto2013}.}
\label{tab:BM_Spectral_Interpretation}
\renewcommand{\arraystretch}{1.5}
\begin{tabularx}{\textwidth}{
  >{\raggedright}p{3.8cm}
  >{\raggedright}X
  >{\raggedright\arraybackslash}X}
\toprule
Parameter
  & Interpretation in stable model
  & Interpretation via BM hitting direction \\
\midrule
$\mu \in S^{d-1}$
  & Mean direction of spectral mass
    concentration
  & Direction of BM drift vector
    $\kappa\mu \in \mathcal{R}^d$ \\
$\kappa \geq 0$
  & Concentration of spectral mass
    around $\mu$
  & Magnitude of drift $\|\kappa  \mu\|$;
    larger $\kappa$ means stronger pull
    toward $\mu$ \\
$\kappa = 0$
  & Uniform spectral measure on $S^{d-1}$
  & Zero drift: symmetric BM with
    uniform hitting direction \\
$\kappa \to \infty$
  & Degenerate point mass at $\mu$
  & Infinite drift: BM hits the sphere
    deterministically at $\mu$ \\
$\bar{R} = I_{d/2}(\kappa)/I_{d/2-1}(\kappa)$
  & Mean resultant length; measures
    degree of anisotropy
    ($\bar{R}=0$: isotropic;
    $\bar{R}\to 1$: degenerate)
  & Expected projection of hitting
    direction onto $\mu$; larger
    $\bar{R}$ means more concentrated
    near $\mu$ \\
\bottomrule
\end{tabularx}
\end{table}

\subsection{Regular Variation Characterisation}
\label{sec:regular_variation}

The connection between the spectral measure $\Gamma$ and the
tail behaviour of $X \sim \mathrm{Stable}(\alpha,\Gamma)$ in
$\mathcal{R}^d$ is made precise by the regular variation
characterisation \citep{Resnick1987, BasrakDavisMikosch2002}.
With $\wG \sim \mathrm{vMF}(\mu,\kappa)$, for any Borel set
$A \subset \Sd$, $\Gamma(\partial A) = 0$, we have
\begin{align}\label{eq:regular_variation}
  \lim_{x \to \infty}
  \frac{P(X/\|X\| \in A,\; \|X\| > x)}
       {P(\|X\| > x)}
  &= \wG(A)
  = \int_A
    f_{\mathrm{vMF}}(s \mid \mu, \kappa)\, ds.
\end{align}
Thus, conditional on lying in the far tail, the probability
that $X$ points into direction set $A$ converges to the
normalised spectral measure evaluated at $A$.  The spectral
measure is therefore the limiting angular distribution of
extreme observations.  For the vMF spectral measure,
\eqref{eq:regular_variation} provides an explicit parametric
form for this limiting distribution, valid in any dimension
$d \geq 2$, controlled by the mean direction $\mu$ and
concentration $\kappa$.  

It is also worth mentioning that the von Mises--Fisher distribution $\mathrm{vMF}(\mu, \kappa)$ on
$S^{d-1}$ is the unique maximum entropy distribution subject to a
fixed mean resultant vector $\mathbb{E}[s] = \bar{s}$
(fixed mean direction $\mu \in S^{d-1}$ and fixed mean resultant
length $\bar{R} = \|\bar{s}\| \in [0,1]$) \citep{MardiaJupp2000}.

\subsection{The Gaussian Boundary}
\label{sec:gaussian_boundary}

At the Gaussian boundary, we show that all spectral measures induces a specific covariance structure. In the case of vMF, the covariance gets a closed form through its second moment matrix.

\begin{theorem}[Gaussian boundary, general $d$]
\label{thm:gaussian_boundary}
Let $X \sim \mathrm{Stable}(\alpha, \Gamma)$ in $\mathcal{R}^d$,
$\alpha \in (0,2)$, with finite spectral measure $\Gamma$ on
$S^{d-1}$ and total mass $\Gamma_\infty = \Gamma(S^{d-1}) < \infty$.
As $\alpha \to 2$,
\begin{equation}
  X \xrightarrow{d} N_d\!\left(0,\, \Sigma_\Gamma\right),
  \qquad
  \Sigma_\Gamma \;=\; 2 \; \Gamma_\infty \int_{S^{d-1}} ss^\top\, \wG(ds).
  \label{eq:gaussian_boundary}
\end{equation}
The limiting covariance $\Sigma_\Gamma$ is determined entirely by
the second moment of the normalised spectral measure
$\widetilde{\Gamma}$ and the total spectral mass $\Gamma_\infty$.
\end{theorem}
 Thus, as per Theorem~\ref{thm:gaussian_boundary}, at the boundary, information on second moment of the spectral measure is retained and all higher-order information is lost.

\begin{corollary}[von Mises--Fisher spectral measure]
\label{cor:vmf_covariance}
Under the conditions of Theorem~\ref{thm:gaussian_boundary},
let $\Gamma = \Gamma_\infty\,\widetilde{\Gamma}$ where
$\widetilde{\Gamma}$ has density $f_{\mathrm{vMF}}(\cdot\,;\mu,\kappa)$
with respect to the uniform measure on $S^{d-1}$,
$\mu \in S^{d-1}$ is the mean direction, and $\kappa \geq 0$
is the concentration parameter. Then
\begin{equation}
  \Sigma_\Gamma
  \;=\; 2\, \Gamma_\infty
  \bigl[\,b(\kappa,d)\,\Id_d + c(\kappa,d)\,\mu\mu^\top\bigr],
\end{equation}
where

\begin{equation}
\begin{aligned}
  a(\kappa,d) &\;=\; \frac{1}{d}
    + \frac{d-1}{d}\,
    \frac{I_{d/2+1}(\kappa)}{I_{d/2-1}(\kappa)},
  \\
  b(\kappa,d) &\;=\; \frac{1 - a(\kappa,d)}{d-1},
  \\
  c(\kappa,d) &\;=\; a(\kappa,d) - b(\kappa,d).
\end{aligned}
\label{eq:abc_kd}
\end{equation}

The matrix $\Sigma_\Gamma$ has eigenvector $\mu$ with eigenvalue
$2\, \Gamma_\infty\,a(\kappa,d)$ and every direction orthogonal to
$\mu$ with eigenvalue $2\,\Gamma_\infty\,b(\kappa,d)$.
The anisotropy ratio $a(\kappa,d)/b(\kappa,d)$ increases
monotonically from $1$ (isotropic, $\kappa=0$, uniform
spectral measure) to $\infty$, which follows from 
\cite{Amos1974}.  

\end{corollary}
\begin{proposition}[Anisotropy under von Mises--Fisher spectral measure]
\label{prop:anisotropy}
Under the conditions of Corollary~\ref{cor:vmf_covariance}, the
anisotropy ratio
\[
  \rho(\kappa,d)
  \;=\;
  \frac{a(\kappa,d)}{b(\kappa,d)}
  \;=\;
  \frac{(d-1)\,a(\kappa,d)}{1-a(\kappa,d)}
\]
satisfies the following:
\begin{enumerate}[label=(\roman*)]
  \item $\rho(0,d)=1$: when $\kappa=0$ the normalised spectral measure $\wG$ is
        uniform and $\Sigma_\Gamma \propto \Id_d$ (isotropic);
  \item $\rho(\kappa,d)$ is strictly increasing in $\kappa$ on
        $(0,\infty)$, so greater spectral concentration along $\mu$
        produces strictly greater anisotropy;
  \item $\lim_{\kappa\to\infty}\rho(\kappa,d)=\infty$: as the
        spectral measure $\wG$ concentrates on $\{\mu\}$ the distribution
        becomes degenerate along $\mu$.
\end{enumerate}
Hence $\kappa$ is a pure anisotropy parameter: it controls the
eigenvalue ratio of the Gaussian limit covariance $\Sigma_\Gamma$
monotonically, independently of $\alpha$ and $\Gamma_\infty$.
\end{proposition}
\begin{remark}
Regular variation equation~\eqref{eq:regular_variation} reveals that the anisotropy
of extreme observations is governed entirely by~$\kappa$: the
limiting angular distribution of tail realisations is
$\mathrm{vMF}(\mu,\kappa)$ for every $\alpha\in(0,2]$.  The anisotropy ratio $\rho(\kappa,d)$
of Proposition~\ref{prop:anisotropy}, derived at the
Gaussian boundary $\alpha=2$, therefore reflects a geometric
property that is in fact exact in the tail at every $\alpha$.
\end{remark}

\begin{corollary}[Bivariate case, $d = 2$]
\label{cor:bivariate}
At $d = 2$, with $\mu = (\cos\eta, \sin\eta)^\top$ and
$R_2(\kappa) = I_2(\kappa)/I_0(\kappa)$,
Corollary~\ref{cor:vmf_covariance} reduces to
\begin{equation}
  \Sigma_\Gamma
  \;=\; \Gamma_\infty
  \!\left(\Id_2 + R_2(\kappa)\,\IA(\eta)\right),
  \qquad
  \IA(\eta) =
  \begin{pmatrix}
    \cos 2\eta & \sin 2\eta \\
    \sin 2\eta & -\cos 2\eta
  \end{pmatrix}.
  \label{eq:BVN_covariance_matrix}
\end{equation}
\end{corollary}

\begin{remark}[Special cases]
\label{rem:special}
\begin{enumerate}[label=(\roman*)]

  \item \emph{Isotropic case} ($\kappa = 0$, uniform spectral
    measure): $a(0,d) = 1/d$, $b(0,d) = 1/d$, $c(0,d) = 0$,
    giving $\Sigma_\Gamma = (2 \, \Gamma_\infty/d)\,\Id_d$. 
    
  \item \emph{Concentrated case} ($\kappa \to \infty$, point
    mass at $\mu$): $a \to 1$, $b \to 0$, $c \to 1$, giving
    $\Sigma_\Gamma \to 2 \,\Gamma_\infty\,\mu\mu^\top$. 
  \item \emph{Generality}: Theorem~\ref{thm:gaussian_boundary}
    holds for any finite spectral measure $\Gamma$ --- discrete,
    continuous, or mixed. 
    
\end{enumerate}
\end{remark}

\section{LePage-type Representation  and Gaussian Truncation Error}
\label{sec:lepage_truncation}

The classical benchmark for representing stable distributions is the LePage series representation \citep{LePage1980,LePage1981} and its generalisation \citep{Rosinski1990}. Let $\{X_i\}_{i=1}^n$ be stable random vectors in $\mathcal{R}^d$.  Let
$s_{ij}  \in\mathcal{S}^{d-1}$, $i=1,\ldots,n$, $j=1,2,\ldots$,
be independent identically distributed draws from the normalised spectral measure $\wG$.  The LePage representation is
\begin{equation}\label{eq:LePage_series}
  X_i = \sum_{j=1}^\infty
  L_{ij}(\alpha)\,s_{ij},
  \quad i=1,\ldots,n,
\end{equation}
where $L_{ij}(\alpha)$ are weights derived from the arrival times of a unit-rate Poisson process on $\mathcal{R}_+$, independent of $\{s_{ij}\}$.

Direct implementation of~\eqref{eq:LePage_series} is computationally demanding; the series is infinite and sorting
Poisson arrival times is expensive. Instead, we use a modified version of the infinitely divisible representation of
\citet{DamienLaudSmith1995}, truncated at $m$ terms, with truncation error captured through a Gaussian component:
\begin{equation}\label{eq:LePage_truncated}
  X_i = \sum_{j=1}^m
  L_{ij}(\alpha)\,s_{ij}
  + \epsilon_i,
  \quad i=1,\ldots,n,
\end{equation}
where $\epsilon_i\sim N_d(0, (\sigma^2/m)\Id_d)$ and $\sigma^2 >0 $ is an unknown parameter that absorbs
the truncation bias.  In the modified representation, $L_{ij}(\alpha)$ is a function of independent Poisson
and inverted-gamma variates.
 
\subsection{A LePage-type representation theorem}

Various authors developed methods to draw from univariate infinitely divisible distributions. We adopt  
 a univariate result by \cite{DamienLaudSmith1995} for stable distributions.

\begin{theorem}[\citet{DamienLaudSmith1995},
  modified for stable distributions]
\label{thm:DLS_bivariate}
  Let $s_1,\ldots,s_m$ be independent
  and identically distributed from  $\vMF(\eta, \kap)$.  Let $r_1,\ldots, r_m$
  be independent and identically
  distributed as inverted gamma with
  parameter $\alpha$ and density
  \begin{equation}\label{eq:IG_density_thm}
    g(r) = \frac{1}{\Gamma(\alpha)}
    r^{-1-\alpha}e^{-1/r},
    \quad r>0.
  \end{equation}
  For $j=1,\ldots,m$, let $y_j$ be
  drawn independently from
  $\mathrm{Poisson}(\lambda_j)$,
  where $\lambda_j= \Gamma_\infty\Gamma(\alpha)
  e^{1/r_j}/m$.  Define
  \begin{align}\label{eq:Xm}
    X^{(m)} &= \sum_{j=1}^m r_j    \left\{y_j - \frac{\lambda_j}     {1+r_j^2}\right\}s_j  
     = \sum_{j=1}^m L_j s_j,
  \end{align}
where $L_j = r_j    \left\{y_j - \lambda_j/( {1+r_j^2})\right\} $.

  Let $\psi_m(t) = \E[e^{i\langle t, X^{(m)}\rangle}]$, $t\in\mathcal{R}^d$,
  denote the characteristic function
  of $X^{(m)}$.  
  Then $\psi_m(t)\to\psi(t)$ as
  $m\to\infty$, where $\psi(t)$ is the
  characteristic function of the
  $\mathrm{Stable}(\alpha,\Gamma)$
  distribution.
\end{theorem}

\begin{figure}[H]
\centering
\includegraphics[width=\textwidth]{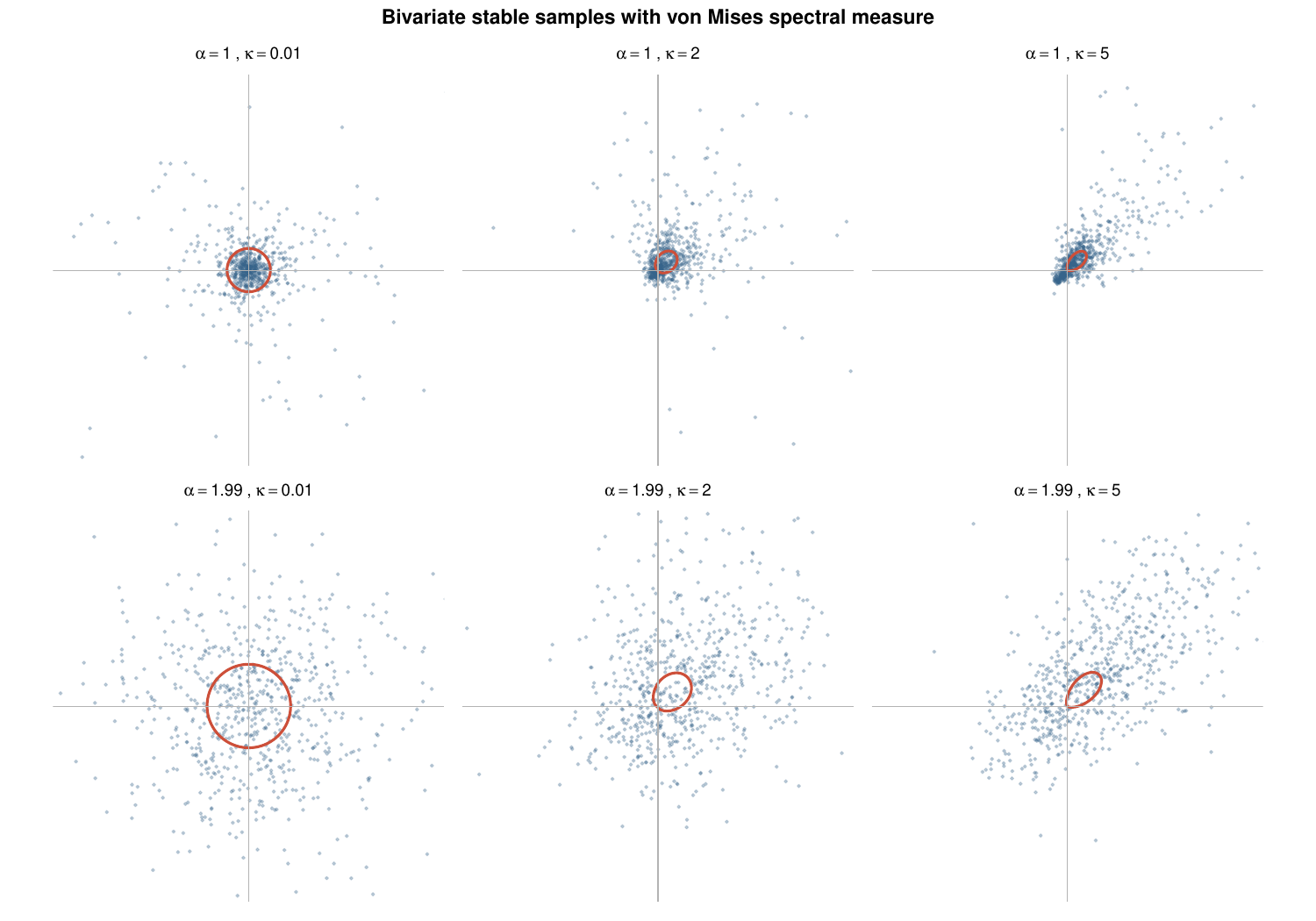}
\caption{Bivariate stable samples generated from the
  LePage construction (Theorem~\ref{thm:DLS_bivariate})
  with von Mises spectral measure $\mathrm{vM}(\mu,\kappa)$,
  $\mu = \pi/4$ and $\Gamma_\infty=1$.  Each panel shows $n = 500$ independent
  realisations of $X_i^{(m)}$ with $m = 200$ terms,
  truncated to $r \in [0.1, 10^6]$.
    The red curve is the polar density
  of the vM spectral measure, scaled for display.
  Rows: stability index $\alpha = 1$ (Cauchy, top)
  and $\alpha = 1.99$ (near-Gaussian, bottom).
  Columns: concentration $\kappa = 0.01$ (near-uniform),
  $\kappa = 2$ (moderate), $\kappa = 5$ (concentrated).
  Points beyond the 95th percentile of the radial
  component $r_{ij}$ are excluded from display.}
  \label{fig:figure1_lepage_samples}

\end{figure}

Figure~\ref{fig:figure1_lepage_samples} illustrates the bivariate
stable distribution generated by the LePage construction
(Theorem~\ref{thm:DLS_bivariate}) with a von Mises spectral measure
$\mathrm{vM}(\mu, \kappa)$.  Each panel shows $n = 500$
independent realisations of $X_i^{(m)}$, with $m = 200$ terms, together with the
polar density of the spectral measure (red curve) centred
at the origin.  The mean direction $\mu = \pi/4$ is fixed
throughout; the concentration parameter $\kappa$ increases
across columns ($\kappa = 0.01, 2, 5$), and the stability
index $\alpha$ varies across rows ($\alpha = 1$, Cauchy;
$\alpha = 1.99$, near-Gaussian).  
The directional structure of the LePage sum is governed
jointly by the vM parameters $(\mu, \kappa)$ through the
spectral measure and by $\alpha$ through the radial
components $r_{ij}$.

\subsection{Construction of a
  Likelihood Function}
\label{sec:construction_likelihood}

Bayesian inference for $\mathrm{Stable}(\alpha,\Gamma)$ requires a
tractable likelihood.  Existing density representations
\citep{Zolotarev1986,AbdulHamidNolan1998} do not factor as a joint
density for $(X, \text{latent variables})$, precluding the
data-augmentation approach that \citet{Buckle1995} used successfully
for univariate stable laws.  We resolve this via a multivariate
extension of \citet{DamienLaudSmith1995}, combined with the
characteristic function of \citet{Kuelbs1973} and the Gaussian
error component in~\eqref{eq:LePage_truncated}, yielding a
likelihood that supports full Bayesian inference.

Let $\{X_1,\ldots,X_n\}$ be independently and identically distributed Stable$(\alpha, \Gamma)$ random vectors. Construct $X^{(m)}_i$ using $r_{ij}, y_{ij}, s_{ij}, j=1, \ldots, m$, as in Theorem~\ref{thm:DLS_bivariate} and define $X^{(m)}_i$ as in equation~\eqref{eq:Xm}, $i=1,\ldots,n$. Define $\epsilon_i=X_i - X^{(m)}_i, \quad i=1,\ldots, n.$ 
Given $r_{ij}, y_{ij}, s_{ij}, j=1, \ldots, m, i=1,\ldots,n$, we assume that $\epsilon_1,\ldots,\epsilon_n$ follow independent  bivariate Gaussian law with mean vector zero and dispersion matrix $(\sigma^2/m) \, \Id_d$; that is,
\begin{eqnarray}
\epsilon_i \sim N_d\!\left(0,\frac{\sigma^2}{m}\Id_d\right), i=1,\ldots,n,
\end{eqnarray}
where $\sigma^2$ is unknown and $\Id_d$ is the identity matrix.

Deriving the likelihood function is now straightforward. 
By conditional independence of the latent variables
$(r_{ij}, y_{ij}, s_{ij})$ given the parameters, the
complete-data likelihood factors as
\begin{align}
  L(\alpha,\eta,\kappa)
  &=
    \left\{\prod_{i=1}^n\prod_{j=1}^m
    \frac{1}{\Gamma(\alpha)}
    \frac{e^{-1/r_{ij}}}
    {r_{ij}^{1+\alpha}}\right\}
    \times
    \left\{\prod_{i=1}^n\prod_{j=1}^m
    \frac{e^{-\lambda_{ij}}
    \lambda_{ij}^{y_{ij}}}
    {y_{ij}!}\right\}
  \nonumber\\
  &\quad\times
    \left\{\prod_{i=1}^n\prod_{j=1}^m
    f_{\mathrm{vMF}}(s_{ij}\mid\eta,\kappa)\right\}
    \times
    \left\{\prod_{i=1}^n
    f(\varepsilon_i\mid\alpha,\eta,\kappa,
    \{r_{ij}\},\{y_{ij}\},
    \{s_{ij}\})\right\}.
  \label{eq:likelihood}
\end{align}
where $\lambda_{ij} = \Gamma_\infty\,\Gamma(\alpha)\,
e^{1/r_{ij}}/m$.

Our approach towards estimating the spectral measure $\Gamma$ is based on the likelihood function derived in equation~\eqref{eq:likelihood}; derivation of the likelihood function is given in Section~\ref{sec:likelihood_derivation} as online material. 
Inference based on approximate likelihood function is not uncommon especially
in the absence of an explicit density function form. For instance, likelihood-based inference in frequency
domain in time series is generally based on the Whittle likelihood which is
the asymptotic distribution of the periodogram; see Choudhuri, Ghosal, and Roy (2003)
and references therein for this example.

Let $\Psi_f = \Gamma_\infty\Gamma(\alpha)$ denote the combined
scale parameter. Based on Theorem~\ref{thm:DLS_bivariate} and
the von Mises -- Fisher spectral density, the complete-data log-likelihood
decomposes as
\begin{equation}\label{eq:log_likli}
  \ell_c(\alpha, \Psi_f, \kappa, \eta)
  = \ell_{\mathrm{Pois}}(\Psi_f)
  + \ell_{\mathrm{Gauss}}(\Psi_f, \kappa, \eta)
  + \ell_{\mathrm{IG}}(\alpha)
  + \ell_{\mathrm{vMF}}(\kappa, \eta).
\end{equation}
Note that $\Gamma_\infty$ enters $\ell_c$ only through $\Psi_f$. For simulation studies, we use specific form of \eqref{eq:log_likli} with $d=2$. In this case we can express $s$ in terms of the directional angle $\theta$. 
With $d=2$, individual components of the log-likelihood function in ~\eqref{eq:log_likli}  are the following.

\paragraph{Poisson term:}
$  \lPois(\Psi_f)
  = \sum_{i,j}\bigl[y_{ij}\log\lam_{ij} - \lam_{ij}
    - \log(y_{ij}!)\bigr],
    \lam_{ij} = \frac{\Psi_f}{m}\exp\!\Bigl(\frac{1}{r_{ij}}\Bigr).
$

\paragraph{Gaussian term:} $  \lGauss
  = - (m/ (2\sigma^2))
    \sum_{i=1}^n
    \bigl\|X_i - X_i^{(m)}\bigr\|^2$, where the series approximation is
$  X_i^{(m)}  = \sum_{j=1}^m    L_{ij}\,s(\theta_{ij})$, 
$  L_{ij} = r_{ij} \left( y_{ij} -    \lam_{ij}/ (1 + r_{ij}^2)\right)$,
 $ \bs(\theta) = \left( \cos\theta \;\; \sin\theta \right) ^\top$.

\paragraph{Inverted-gamma on $r_{ij}$:}
$  \lIG(\alpha)
  = -nm\log C_\delta(\alpha)
    -(1+\alpha)\sum_{i,j}\log r_{ij}
    - \sum_{i,j} (1/r_{ij}).$

\paragraph{Von Mises on $\theta_{ij}$:}
$  \lvM(\kap, \eta)
  = -nm\log I_0(\kap)
    + \kap\sum_{i,j}\cos(\theta_{ij} - \eta).
$

\begin{remark}[Smoothness of the likelihood in $\alpha$]
\label{rem:smooth_lik}
The complete-data log-likelihood $\ell_c$ in
\eqref{eq:log_likli} is $C^\infty$ in $\alpha \in (0,2)$,
including at the Cauchy boundary $\alpha = 1$.
The Poisson component does not depend on $\alpha$ after reparametrisation through $\Psi_f$.  vM terms do not depend on $\alpha$.
The IG term $\ell_{\mathrm{IG}}(\alpha)$ is $C^\infty$
in $\alpha$ since $\Gamma(\alpha)$ is $C^\infty$ on $(0,2)$.
The Gaussian term depends on $\alpha$ only through
$\Gamma(\alpha)$ in $\lambda_{ij} = \Psi_f e^{1/r_{ij}}/m$,
which is likewise $C^\infty$.
The practical consequence is that the MCMC sampler for
$\alpha$ requires no special treatment at $\alpha = 1$
and the chain crosses this boundary freely. A stronger result — $C^\infty$ smoothness of the characteristic function itself — is established in Theorem~A1 of the Appendix.
\end{remark}

\subsection{A truncated version}

Let $0<\rmin<\rmax<\infty$ be the lower and upper
truncation thresholds for the radial jump sizes.
The truncated inverted-gamma distribution
$\IG(\alpha,\rmin,\rmax)$ has density
\begin{equation}\label{eq:IG_density}
  g(r \mid \alpha,\rmin,\rmax)
  = \frac{r^{-\alpha-1}e^{-1/r}}
    {\Cab(\alpha)},
  \quad r\in(\rmin,\rmax),
\end{equation}
where the normalising constant is given by
\begin{equation}\label{eq:C_norm}
  \Cab(\alpha)
  = \int_{\rmin}^{\rmax}
    r^{-\alpha-1}e^{-1/r}\,dr
  = \Gamma(\alpha)\bigl[P(\alpha,1/\rmin)
    - P(\alpha,1/\rmax)\bigr],
\end{equation}
with $P(\alpha,x)=\int_0^x t^{\alpha-1}e^{-t}\,dt /\Gamma(\alpha)$,
the lower incomplete gamma function.

The truncated radial density at \eqref{eq:IG_density} is a computational regularisation of the
full inverted gamma density \eqref{eq:IG_density_thm}: the lower
truncation at $r_{\min} > 0$ removes radial components that are
negligibly small in magnitude, while the upper truncation at $r_{\max} < \infty$ guards
against numerical overflow from the rare but large radial
components in the upper tail of the inverted gamma density.
As $r_{\min} \to 0$ and
$r_{\max} \to \infty$, the normalising constant $C_{r_{\min}}(\alpha)
\to \Gamma(\alpha)$  and the characteristic
function of the truncated LePage sum converges to that of the
untruncated sum (Proposition~\ref{prop:truncation} in the Appendix).
The stable limit $\psi(t)$ of Theorem~\ref{thm:DLS_bivariate}
is therefore recovered in the joint limit, confirming that
truncation introduces no asymptotic bias. 
In practice, $r_{\min}$ and $r_{\max}$ are set as functions of
the sample size $m_{\rm gen}$. We use $r_{\min} = 1/\log (m_{\rm gen})$
and $r_{\max} = m_{\rm gen}^{1/\hat\alpha}$, where $\hat\alpha$
is the Koutrouvelis initialisation estimate of $\alpha$
(Section~\ref{sec:initialisation}). Nuances relating to $m_{\rm gen}$ are discussed in Section~\ref{sec:adaptive}.  
\subsection{The Psi Reparametrization}

For computational convenience,
$\Gamma_\infty$ is reparametrized as
\begin{equation}\label{eq:Psi_def}
  \Psi = \Gamma_\infty\Cab(\alpha)
  = \frac{\text{mean Poisson intensity per term}}
    {e^{1/r}/m}.
\end{equation}
We can easily shift between $\Gamma_\infty$ and $\Psi$ through the conversion formula as  
  $\Gamma_\infty = \Psi/\Cab(\alpha)$.

\begin{remark}[Notation]
  Throughout, $\Psi_f=\Gamma_\infty
  \Gamma(\alpha)$ refers to the
  Poisson intensity of the full
  untruncated series
  (Theorem~\ref{thm:DLS_bivariate}),
  while $\Psi=\Gamma_\infty\Cab(\alpha)$
  refers to the truncated version
  used in the likelihood.
  Implementation uses $\Psi$.
\end{remark}

\begin{remark}[Why $\Psi$ is natural]
  The parameter $\Psi$ appears directly in the
  Poisson intensities $\lambda_{ij}=(\Psi/m)
  e^{1/r_{ij}}$ and is identified by
  $\ell_{\rm Pois}(\Psi)$ independently of
  $\Cab(\alpha)$.  The reparametrization from
  $\Gamma_\infty$ to $\Psi$ avoids the need to
  compute $\Cab(\alpha)$ in the Poisson
  likelihood update, making the Gibbs step
  for $\Psi$ a standard log-normal slice
  sampler.
\end{remark}

\subsection{Anatomy of construction}

For $d=2$, our approximate series representation is
\begin{equation}\label{eq:DSL_truncated_series}
  X_i^{(m)}
  = \sum_{j=1}^m L_{ij}\,s(\theta_{ij}),
 \end{equation}
where $r_{ij}\sim\mathrm{IG}(\alpha)$,
$y_{ij}\sim\mathrm{Poisson}(\lambda_{ij})$,
$\theta_{ij}\sim\vM(\eta,\kap)$, and
$\lambda_{ij} = (\Psi/m)\exp(1/r_{ij})$.

The first term $r_{ij} y_{ij}$ is the raw jump: the
product of the radial scale $r_j$ and the discretised
Poisson count $y_j$.  The second term removes the
expectation to ensure the series is centred, analogous
to the centering in the classical LePage series for
$\alpha\geq 1$. This is a truncated LePage series with
four key modifications.

\begin{enumerate}[label=(\roman*)]
\item \textbf{Radial component $r_{ij}$:}
In the classical LePage series,
the radial jumps are
$\Gamma_j^{-1/\alpha}$, where
$\Gamma_j=\gamma_1+\cdots+\gamma_j$
are the arrival times of a
unit-rate Poisson process
($\gamma_i\overset{\rm iid}{\sim}
\mathrm{Exp}(1)$).  In our
representation, $r_{ij}$ plays
the role of $\Gamma_j^{-1/\alpha}$;
its marginal distribution is the
truncated inverted-gamma density at \eqref{eq:IG_density}.

\item \textbf{Poisson bin counts $y_{ij}$:}
The classical LePage series sums
over all Poisson arrivals
individually.  Theorem~\ref{thm:DLS_bivariate}
partitions the intensity measure
into $m$ bins of equal width $1/m$,
and $y_{ij}\sim\mathrm{Poisson}
(\lambda_{ij})$ counts the arrivals
in the $j$-th bin for observation
$i$.  The expectation $\lambda_{ij}
=(\Psi/m)e^{1/r_{ij}}$ gives the
expected number of arrivals in that
bin, connecting the bin width $1/m$
to the total jump intensity $\Psi$.

  \item \textbf{Centering $\lambda_{ij}/(1+r_{ij}^2)$:}
    The classical LePage series requires centering for
    $\alpha\geq 1$ to ensure convergence.  The term
    $-r_{ij}\lambda_{ij}/(1+r_{ij}^2)$ provides this
    centering. Each term $L_{ij}\,s(\theta_{ij})$ in equation~\eqref{eq:DSL_truncated_series}
is a centred directional jump in $\R^2$.

\item \textbf{The direction $s(\theta_{ij})$:}
$s(\theta_{ij}) = (\cos\theta_{ij}, \sin\theta_{ij})^\top$ is a
unit vector on $\mathcal{S}^1$ drawn from the spectral
measure $\vM(\eta,\kap)$.  It represents the
tail direction of the $j$-th jump --- the direction
in which the $j$-th component pulls the sum.
  \end{enumerate}

\paragraph{Geometric interpretation:}
Each term $L_{ij}\,s(\theta_{ij})$ is a random vector
in $\R^2$ pointing in direction $\theta_{ij}$ with signed
magnitude $L_{ij}$.  The sum $\sum_{j=1}^m L_{ij}\,s(\theta_{ij})$
accumulates these directional jumps, and by the
convergence of the LePage series, the distribution of
this sum approaches the bivariate stable distribution
$\mathrm{Stable}(\alpha,\Gamma)$ as $m\to\infty$. The Gaussian approximation error term
$\varepsilon_i \sim N_2(0, (\sigma^2/m) \Id_2)$ captures
the contribution of the unrepresented tail beyond $m$ terms.

\section{$\alpha$ Identification}
\label{sec:alpha_identification}

From the inverted-gamma component of the likelihood function, the score for
$\alpha$ is
\begin{equation}
  \frac{\partial\lc}{\partial\alpha}
  = \frac{\partial\lIG}{\partial\alpha}
  = -nm\frac{\Cab'(\alpha)}{\Cab(\alpha)}
    - \sum_{i,j}\log r_{ij}.
\end{equation}
The sufficient statistic for $\alpha$
is $\sum_{i,j}\log r_{ij}$, identified
from the radial draws $\{r_{ij}\}$
independently of $(\kap,\eta,\Psi)$.
The shape of the truncated
$\IG(\alpha,\rmin,\rmax)$ distribution
changes with $\alpha$, and this shape
change is precisely what makes
$\alpha$ identifiable from
$\{r_{ij}\}$, as quantified by the
strictly positive Fisher information
in Proposition~\ref{prop:alpha_id}.

\begin{proposition}[$\alpha$ identified for all $\alpha\in(0,2)$]
\label{prop:alpha_id}
The complete-data Fisher information for $\alpha$ from the
radial draws $\{r_{ij}\}$ is
\begin{equation}\label{eq:FI_alpha}
  \mathcal{F}_\alpha
  = nm\left[\frac{C_{r_{\min},r_{\max}}''(\alpha)}
             {C_{r_{\min},r_{\max}}(\alpha)}
    -\left(\frac{C_{r_{\min},r_{\max}}'(\alpha)}
               {C_{r_{\min},r_{\max}}(\alpha)}\right)^2
    \right] > 0,
  \qquad \alpha\in(0,2).
\end{equation}
The strict positivity follows from the exponential family
structure of the truncated $\mathrm{IG}(\alpha,r_{\min},r_{\max})$
distribution (Corollary~\ref{cor:fisher}, Appendix).
Table~\ref{tab:FI} reports $\mathcal{F}_\alpha/nm$ at
selected values of $\alpha$, with $r_{\min}=0.1$ and
$r_{\max}=10^6$.   
\end{proposition}
\begin{table}[H]
\centering
\caption{Complete-data Fisher information per observation
  $\mathcal{F}_\alpha/nm = \mathrm{Var}_\alpha[\log r]$
  at selected $\alpha$, with $r_{\min}=0.1$, $r_{\max}=10^6$.}
\label{tab:FI}
\begin{tabular}{cc}
\toprule
$\alpha$ & $\mathcal{F}_\alpha / nm$ \\
\midrule
$0.25$ & $244.949$ \\
$0.75$ & $4.611$ \\
$1.00$ & $1.490$ \\
$1.50$ & $0.265$ \\
$1.75$ & $0.125$ \\
$1.90$ & $0.079$ \\
$1.95$ & $0.068$ \\
$1.99$ & $0.060$ \\
\bottomrule
\end{tabular}
\end{table}

\begin{remark}[Near-Gaussian identifiability]
Table~\ref{tab:FI} shows that $\mathcal{F}_\alpha/nm$
decreases from $244.949$ at $\alpha=0.25$ to $0.060$
at $\alpha=1.99$, with the steepest drop between
$\alpha=1.0$ and $\alpha=1.5$.  Beyond $\alpha=1.75$
the values cluster near zero, consistent with
Theorem~\ref{thm:gaussian_boundary}: as $\alpha\to 2$
the stable law converges to a Gaussian, and $\alpha$
becomes difficult to identify from the radial components
alone.  Simulation confirms wider posterior intervals
for $\alpha \geq 1.75$. 

The complete-data Fisher information $\mathcal{F}_\alpha$
is decreasing in $\alpha$ (Table~\ref{tab:FI}), but the
effective information reaching the sampler is not monotone:
at small $\alpha$, the large radial components that carry
most information about $\alpha$ lie beyond $r_{\max}$ and
are not directly observed, so the sampler must impute them
from a truncated distribution.  
\end{remark}

\section{Initialisation and diagnostics}
\label{sec:initialisation_diagnostics}
\subsection{Initialisation}
\label{sec:initialisation}

The initial value $\hat\alpha_0$ is obtained by applying the characteristic function regression of \citet{Koutrouvelis1980} to four univariate projections of the data (at $0^\circ$, $45^\circ$, $90^\circ$, $135^\circ$), taking the median of valid estimates. The Koutrouvelis regression linearises $\log(-\log|\hat\phi_n(t)|^2)$ in $\log|t|$ to read off $\alpha$ as a slope. This performs well when the data are genuinely heavy-tailed, so that the empirical characteristic function decays at a rate clearly distinct from the Gaussian. Near the boundary $\alpha \to 2$, however, the slope flattens and the estimate becomes unreliable, coinciding with a collapse of the Fisher information for $\alpha$; see Table~\ref{tab:FI}.  We therefore supplement the Koutrouvelis step with a near-normality check: if the sample excess kurtosis of the pooled projections falls below $1.0$ --- a threshold inconsistent with $\alpha < 1.8$ at $n = 2000$ --- the regression is bypassed and $\hat\alpha_0 = 1.95$ is used instead. All other parameters are initialised from the same projections using moment-based estimates.

The spectral parameters $(\eta,\kap)$
are initialised from the unit-vector
projections $\bu_i=X_i/\|X_i\|$. The
mean direction $\hat\eta_0$ is the
sample circular mean and the
concentration $\hat\kap_0$ is
obtained via the piecewise
\citet{BestFisher1981} approximation
to $A^{-1}(\hat{\bar{R}}_0)$, where
$\hat{\bar{R}}_0$ is the sample mean
resultant length.  The total mass
$\Gamma_\infty$ is initialised by applying
the Koutrouvelis scale estimator to
each of  eight equally spaced
projection directions, computing
$\hat\Gamma_\infty^{(k)}=(\hat\gamma^{(k)})
^{\hat\alpha_0}/\hat\II(\bu_k;
\hat\alpha_0,\hat\eta_0,\hat\kap_0)$
per direction, and taking the median
of valid candidates.  Full details
of all initialisations are given in
Appendix~\ref{app:MCMC_initialisation}.

\subsection{Convergence Diagnostics}
\label{sec:diagnostics}

We monitor convergence using two complementary diagnostics computed at
each iteration.

\noindent\textbf{SSE ratio.}
The residual sum of squares
$\mathrm{SSE}=\sum_i\|X_i-X_i^{(m)}\|^2$,
normalised by $2n\hat\sigma^2$, should equal $1$ at convergence,
since $\sigma^2$ is updated by a conjugate inverse-gamma step that
absorbs the current residual variance.
A value near $1$ confirms internal self-consistency of the Gaussian
approximation, but does not guarantee convergence of
$(\alpha,\Psi,\eta,\kappa)$: if these are poorly estimated,
$\sigma^2$ simply inflates to compensate.

\noindent\textbf{logr gap.}
The sufficient statistic for $\alpha$ under the truncated
$\mathrm{IG}(\alpha)$ distribution is $\overline{\log r}=
\sum_{i,j}\log r_{ij}/(nm)$.
At stationarity this should agree with its expected value
$\mathrm{ref\_logr}(\alpha) =
E_{\mathrm{IG}(\alpha,\delta,r_{\mathrm{upper}})}[\log r]$
under the current draw of $\alpha$.
We monitor the gap
\begin{equation}
  \mathrm{logr\_gap}
  = \overline{\log r} - \mathrm{ref\_logr}(\alpha),
\end{equation}
which serves as a real-time diagnostic for $\alpha$ convergence;
a persistent non-zero gap indicates that the chain has not found
the correct stability index; also see Theorem~\ref{thm:sign} in the online material.
Simulation showed that $\alpha$ convergence does not automatically
imply convergence of $\kappa$. We use $n=2000$
observations with a chain of $50{,}000$ iterations and a burn-in
of $20{,}000$ to ensure adequate mixing across all parameters.
Correct convergence requires both SSE ratio $\approx 1$ and
logr\_gap $\approx 0$ simultaneously; SSE ratio $\approx 1$ with
logr\_gap $\neq 0$ signals that $\sigma^2$ has inflated to absorb
model misfit while $\alpha$ remains biased.

\subsection{Prior Distributions}

All priors are uniform, reflecting
broad uncertainty while excluding
degenerate values:
\begin{align*}
  \alpha &\sim \mathrm{Uniform}
    (10^{-4},\, 1.999),\\
  \eta &\sim \mathrm{Uniform}
    (0,\, 2\pi),\\
  \kap &\sim \mathrm{Uniform}
    (0.1,\, 50),\\
  \Gamma_\infty &\sim \mathrm{Uniform}
    (0.1,\, 50).
\end{align*}
The upper bound $\alpha_{\rm upper}
=1.999$ excludes the Gaussian
boundary; the lower bound
$\alpha_{\rm lower}=10^{-4}$
excludes numerical degeneracy.
The reparametrized mass
$\Psi=\Gamma_\infty\Cab(\alpha)$
inherits a uniform prior on
$(0.1,50)$.  The approximation
variance $\sigma^2$ is assigned
an informative inverse-gamma
prior with mean near unity,
reflecting the expectation that
the Gaussian truncation error
is of moderate size.

\section{Simulation Study}
\label{sec:simulation}

\subsection{Purpose}

The simulation study has two objectives.
First, to assess recovery of the stability
index $\alpha$ across the full range
$(0,2)$, including the boundaries
$\alpha\to 0$ and $\alpha\to 2$.
Second, misspecification with respect to $m$ and $\rmax$.

\subsection{Adaptive Truncation
  Parameters}
\label{sec:adaptive}

The LePage series requires a lower
truncation bound $\rmin$ (minimum
jump size) and an upper bound
$r_{\rmax}$ (maximum jump size).
The series structure changes across
$\alpha$. Near $\alpha=0$, the
distribution is dominated by a few
very large jumps requiring a large
$r_{\rmax}$. Near $\alpha=2$,
the Gaussian structure arises from
accumulation of many small jumps
requiring a large number of terms.
For data generation, we set
$m_{\rm gen}$ adaptively.
\begin{equation}\label{eq:mgen}
  m_{\rm gen}
  = \max\  \left(30,\min\!\left(10^4,
    \left\lfloor 30 + \frac{5}{\alpha^2}
    + \frac{15}{(2-\alpha)^2}
    \right\rfloor\right)\right),
\end{equation}
with truncation bounds
$  \rmin  = \max\  \left(0.1,    1/ \ln(m_{\rm gen}) \right)$,  $  \rmax
  = \min\!\left(m_{\rm gen}^{1/\alpha},
    10^6\right)$.
The lower bound $\rmin$ prevents
numerical overflow in the Poisson
intensity $\lambda_{ij}=(\Psi/m)
e^{1/r_{ij}}$. The adaptive value
$\rmin=1/\ln(m_{\rm gen})$ is the
smallest for which $e^{1/\rmin}
\approx m_{\rm gen}$ remains within
double precision.  
The upper bound $r_{\max}=10^6$ with lower bound $\rmin=0.1$
retains $98.97\%$ of the $\mathrm{IG}(0.25)$ mass at $\alpha=0.25$,
the most demanding case; for $\alpha \geq 0.5$ the retained mass
exceeds $99.9\%$. The small truncation loss at $\alpha=0.25$
induces a minor positive bias in $\widehat{\Gamma_\infty}$, consistent
with the simulation results in Table~\ref{tab:sim}.

\begin{remark}[$m_{\rm gen}$
  versus $m$]
  $m_{\rm gen}$ is used solely for
  data generation and varies with
  $\alpha$.  The MCMC 
  uses a fixed $m=30$ terms
  for all the runs, with
  $\rmin$ and $r_{\rmax}$
  fixed at the values corresponding
  to the true $\alpha$.
\end{remark}

\subsection{Simulation Designs}

\noindent\textbf{Main simulation.}
LePage series data with $n=2000$
observations, $m=30$ terms, across
$\alpha\in\{0.25,0.50,0.75,1.00,
1.25,1.50,1.75,1.99\}$, with
$\Gamma_\infty=1$, $\eta=\pi/2$, $\kap=2$,
$\sigma^2=1$.  Ten replications per
$\alpha$ value, each
with 50{,}000 MCMC iterations,
20{,}000 burn-in, thinning by 5,
giving 6{,}000 post-burnin samples.

\noindent\textbf{BVN comparison.}
At $\alpha=1.99$, ten additional runs
use exact bivariate Gaussian data
$X_i\sim N_2({0},\Sigma_\Gamma)$ with
$\Sigma_\Gamma=\Gamma_\infty(\Id_2+R_2(\kap)A(\eta))$,
bypassing the LePage series entirely.
This comparison isolates inference algorithm
performance from data generation
approximation error.

\subsection{Simulation Results}
\label{sec:main_results}

Table~\ref{tab:sim} summarises posterior performance across the
$\alpha$ grid ($n=2000$, $m=30$, 10 replications).

\textbf{Stability index $\alpha$.}
For $\alpha \in [0.75, 1.75]$ the bias satisfies
$|\widehat\alpha - \alpha| \leq 0.036$ and coverage ranges from
$0.5$ to $0.9$, consistent with the moderate Fisher information
in this region (Table~\ref{tab:FI}).
At $\alpha = 0.25$ there is a small upward bias of $+0.036$;
at $\alpha = 1.99$ the downward bias reaches $-0.105$, reflecting
the near-zero Fisher information near the Gaussian boundary
(Section~\ref{sec:alpha_identification}).

\textbf{Mean direction $\eta$.}
The mean direction is the best-recovered parameter: bias
$|\widehat\eta - \eta_0| < 0.022$ for all $\alpha \in [0.25,1.99]$,
confirming that the von Mises mean direction is identifiable
independently of the stability index.

\textbf{Concentration $\kappa$.}
For $\alpha \leq 0.50$ the posterior mean of $\kappa$ is
substantially below the true value ($\widehat\kappa \approx 1.6$
vs.\ $\kappa_0 = 2$).
This reflects the difficulty of identifying the spectral
concentration when the distribution is dominated by a few very
large jumps.
For $\alpha \in [0.75, 1.99]$ estimation improves markedly,
with $|\mathrm{Bias}_\kappa| \leq 0.079$.

\textbf{Mass parameter $\Gamma_\infty$.}
For $\alpha \in [0.75, 1.75]$ the bias in $\widehat{ \Gamma_\infty}$ is
negligible ($|\mathrm{Bias}_{\Gamma_\infty}| \leq 0.025$).
At $\alpha = 0.25$ there is a systematic upward bias of $+0.831$,
consistent across all ten replications (range $+0.43$ to $+1.34$).
As shown in Section~\ref{sec:adaptive}, the upper truncation bound
$r_{\max} = 10^6$ retains $98.97\%$ of the $\mathrm{IG}(0.25)$
mass; the residual truncation loss causes the sampler to
overestimate $\Psi = \Gamma_\infty\,\Gamma(\alpha)$ to compensate.
This bias attenuates rapidly and is negligible by $\alpha = 0.75$.

\begin{table}[t]
\centering
\caption{Simulation results for the bivariate stable--von Mises model
  ($n=2000$, $m=30$, $\kappa_0=2$, $\Gamma_{\infty,0}=1$,
  $\eta_0=\pi/4$; 10 replications).
  Bias = posterior mean minus truth (circular wrapping for $\eta$).
  Row marked ${}^{\kappa_4}$ use the kurtosis-based initialisation
  (Section~\ref{sec:initialisation}); all other rows use Koutrouvelis ECF
  initialisation.
  Row \textsc{bvn}: bivariate normal data with covariance matrix at \eqref{eq:BVN_covariance_matrix}. }
\label{tab:sim}
\small
\setlength{\tabcolsep}{5pt}
\begin{tabular}{c rr c rr c rr c rr}
\toprule
 & \multicolumn{2}{c}{$\alpha$}
 &
 & \multicolumn{2}{c}{$\eta$}
 &
 & \multicolumn{2}{c}{$\kappa$}
 &
 & \multicolumn{2}{c}{$\Gamma_\infty$} \\
\cmidrule(lr){2-3}\cmidrule(lr){5-6}\cmidrule(lr){8-9}\cmidrule(lr){11-12}
True $\alpha$
  & $\widehat\alpha$ & Bias$_\alpha$
  &
  & $\widehat\eta$  & Bias$_\eta$
  &
  & $\widehat\kappa$ & Bias$_\kappa$
  &
  & $\widehat \Gamma_\infty$ & Bias$_{\Gamma_\infty}$ \\
\midrule
$0.25$ & $0.286$ & $+0.036$ && $1.564$ & $-0.007$ && $1.605$ & $-0.395$ && $1.831$ & $+0.831$ \\
$0.50$ & $0.467$ & $-0.033$ && $1.568$ & $-0.003$ && $1.707$ & $-0.293$ && $1.084$ & $+0.084$ \\
$0.75$ & $0.714$ & $-0.036$ && $1.572$ & $+0.001$ && $1.984$ & $-0.016$ && $1.003$ & $+0.003$ \\
$1.00$ & $0.971$ & $-0.029$ && $1.569$ & $-0.002$ && $1.990$ & $-0.010$ && $0.996$ & $-0.004$ \\
$1.25$ & $1.226$ & $-0.024$ && $1.567$ & $-0.004$ && $2.048$ & $+0.048$ && $0.976$ & $-0.024$ \\
$1.50$ & $1.497$ & $-0.004$ && $1.567$ & $-0.004$ && $2.069$ & $+0.069$ && $0.975$ & $-0.025$ \\
$1.75$ & $1.730$ & $-0.020$ && $1.576$ & $+0.005$ && $2.079$ & $+0.079$ && $0.988$ & $-0.012$ \\
$1.99$    & $1.886$ & $-0.105$ && $1.550$ & $-0.021$ && $1.955$ & $-0.046$ && $1.067$ & $+0.067$ \\
\addlinespace
\textsc{bvn}
          & $1.940$ & $-0.060$ && $3.674$ & $+0.847$ && $0.184$ & $-1.816$ && $0.474$ & $-0.526$ \\
\textsc{bvn}$^{\kappa_4}$
          & $1.993$ & $+0.003$ && $3.087$ & $+1.516$ && $0.098$ & $-1.902$ && $1.004$ & $+0.004$ \\
\bottomrule
\end{tabular}
\par\smallskip
\raggedright\footnotesize
${}^{\kappa_4}$ If the sample excess kurtosis of the pooled
projections falls below $1.0$, the Koutrouvelis regression is bypassed
and $\hat\alpha_0=1.95$ is used.
For the \textsc{bvn} row without this guard, one replication received fallback option with 
$\hat\alpha_0=0.70$ and failed to converge ($\widehat\alpha=1.530$ after
50{,}000 iterations); the guard eliminates this failure, giving
$\widehat\alpha=1.993$.
\end{table}

\begin{table}[t]
\centering
\caption{Sensitivity of posterior means to the number of LePage terms $m$
  and the radial truncation $r_{\max}$
  ($n=2000$, $\kappa_0=2$, $\Gamma_{\infty,0}=1$, $\eta_0=\pi/2$; 10 replications).
  Bias $=$ posterior mean $-$ truth.
  The standard setting used in Table~\ref{tab:sim} is $m=30$,
  $r_{\max}=10^6$ (marked $\dagger$).}
\label{tab:sensitivity}
\footnotesize
\setlength{\tabcolsep}{3.5pt}
\begin{tabular}{cc rr c rr c rr c rr}
\toprule
 & &
  \multicolumn{2}{c}{$\alpha$} &
  &
  \multicolumn{2}{c}{$\eta$} &
  &
  \multicolumn{2}{c}{$\kappa$} &
  &
  \multicolumn{2}{c}{$z_\infty$} \\
\cmidrule(lr){3-4}\cmidrule(lr){6-7}\cmidrule(lr){9-10}\cmidrule(lr){12-13}
$m$ & $r_{\max}$
  & $\widehat\alpha$ & Bias$_\alpha$
  &
  & $\widehat\eta$ & Bias$_\eta$
  &
  & $\widehat\kappa$ & Bias$_\kappa$
  &
  & $\widehat z_\infty$ & Bias$_{z_\infty}$ \\
\midrule
\multicolumn{13}{l}{\textit{True $\alpha = 0.25$}} \\[2pt]
$30$ & $10^4$          & $0.352$ & $+0.102$ && $1.566$ & $-0.005$ && $1.733$ & $-0.267$ && $2.454$ & $+1.454$ \\
$30$ & $10^6$$^\dagger$& $0.286$ & $+0.036$ && $1.564$ & $-0.007$ && $1.605$ & $-0.395$ && $1.831$ & $+0.831$ \\
$30$ & $10^8$          & $0.264$ & $+0.014$ && $1.565$ & $-0.006$ && $1.509$ & $-0.491$ && $1.664$ & $+0.664$ \\[3pt]
$50$ & $10^4$          & $0.342$ & $+0.092$ && $1.566$ & $-0.005$ && $1.759$ & $-0.241$ && $2.312$ & $+1.312$ \\
$50$ & $10^6$          & $0.279$ & $+0.029$ && $1.560$ & $-0.011$ && $1.634$ & $-0.366$ && $1.754$ & $+0.754$ \\
$50$ & $10^8$          & $0.260$ & $+0.010$ && $1.563$ & $-0.008$ && $1.541$ & $-0.459$ && $1.600$ & $+0.600$ \\
\midrule
\multicolumn{13}{l}{\textit{True $\alpha = 1.99$}} \\[2pt]
$30$ & $10^4$          & $1.819$ & $-0.171$ && $1.550$ & $-0.021$ && $1.874$ & $-0.126$ && $1.112$ & $+0.112$ \\
$30$ & $10^6$$^\dagger$& $1.886$ & $-0.105$ && $1.550$ & $-0.021$ && $1.955$ & $-0.046$ && $1.067$ & $+0.067$ \\
$30$ & $10^8$          & $1.814$ & $-0.176$ && $1.552$ & $-0.019$ && $1.924$ & $-0.076$ && $1.113$ & $+0.113$ \\[3pt]
$50$ & $10^4$          & $1.839$ & $-0.151$ && $1.553$ & $-0.018$ && $1.896$ & $-0.104$ && $1.079$ & $+0.079$ \\
$50$ & $10^6$          & $1.848$ & $-0.142$ && $1.554$ & $-0.017$ && $1.892$ & $-0.108$ && $1.079$ & $+0.079$ \\
$50$ & $10^8$          & $1.848$ & $-0.142$ && $1.550$ & $-0.021$ && $1.818$ & $-0.182$ && $1.058$ & $+0.058$ \\
\bottomrule
\end{tabular}
\end{table}

\textbf{Near-Gaussian boundary check (\textsc{BVN}).}
As a boundary check, we apply the sampler to data generated from
a bivariate normal distribution ($\alpha = 1.99$, $\kappa_0 = 2$).
With ECF based initialisation for $\alpha$, the posterior mean $\widehat\alpha = 1.940$ is close to $1.99$
with coverage $0.9$, confirming that the sampler does not
catastrophically misidentify Gaussian data; kurtosis based initialisation recovers alpha. 
However, $\kappa$ is severely under-estimated
($\widehat\kappa = 0.184$, $\mathrm{Bias}_\kappa = -1.816$)
and $\Gamma_\infty$ is biased downward ($\widehat{ \Gamma_\infty} = 0.474$).
The large bias in $\widehat\eta$ ($+0.847$) is expected:
when $\kappa \approx 0$ the von Mises distribution is nearly
uniform and $\eta$ is unidentifiable.

\textbf{Sensitivity with respect to $m$ and $\rmax$.}
Table~\ref{tab:sensitivity} reports posterior means and biases for
$\alpha=0.25$ and $\alpha=1.99$ across six combinations of
$m\in\{30,50\}$ and $r_{\max}\in\{10^4,10^6,10^8\}$.
The directional parameter $\eta$ is essentially insensitive to both
tuning choices, with biases no larger than $0.021$ across all twelve
cells.
For $\alpha$ and $\kappa$, increasing $r_{\max}$ from $10^4$ to $10^6$
yields a noticeable reduction in bias at $\alpha=0.25$: the bias in
$\widehat\alpha$ falls from $+0.102$ to $+0.036$ and
$\widehat{\Gamma_\infty}$ improves substantially, reflecting the importance
of an adequate radial truncation for heavy-tailed data.
Further increasing $r_{\max}$ to $10^8$ gives only marginal additional
improvement.
At $\alpha=1.99$ the dependence on $r_{\max}$ is weaker and
non-monotone, consistent with the lighter tails near the Gaussian
boundary.
Increasing $m$ from $30$ to $50$ produces small and inconsistent
changes throughout: biases shift by at most $0.010$ in either
direction with no systematic gain.
The standard setting $m=30$, $r_{\max}=10^6$ therefore offers a
reasonable balance between accuracy and computational cost and is used
throughout the main analysis.

\begin{figure}[t]
\centering
\includegraphics[width=0.85\textwidth]{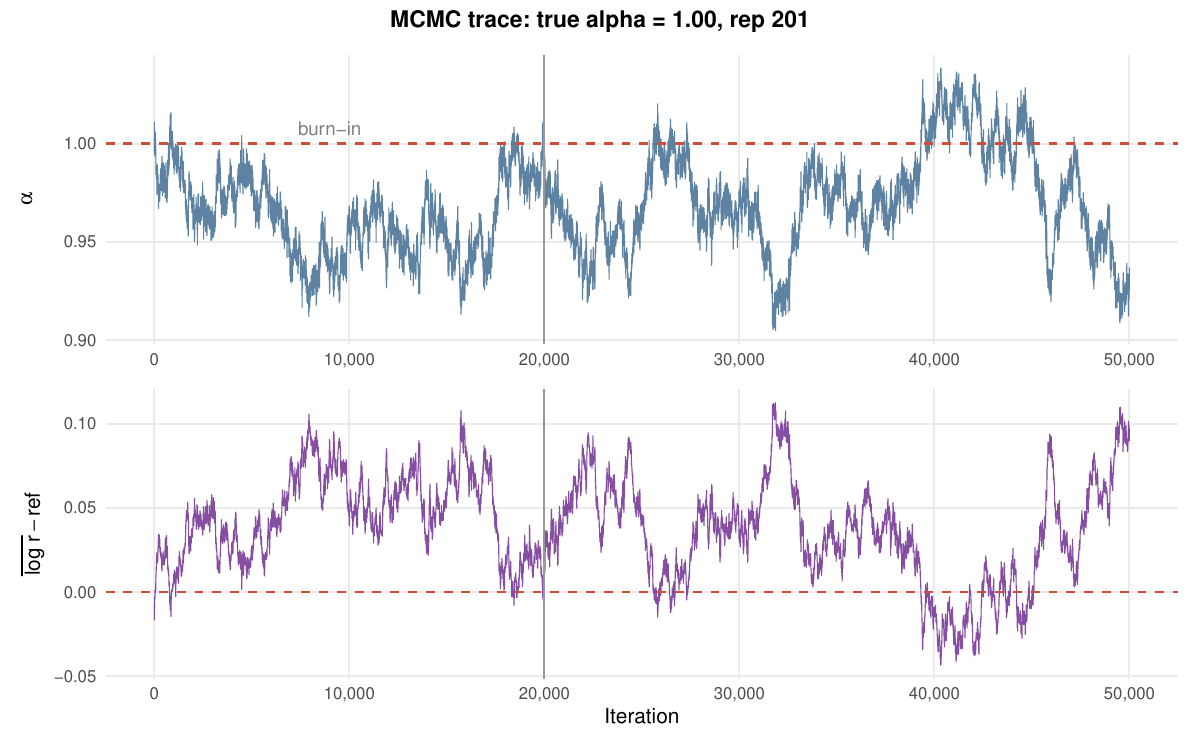}
\caption{MCMC trace for replicate~201
  (true $\alpha = 1.00$, $n = 2{,}000$, $m = 30$).
  \textit{Top:} sampled $\alpha$; dashed line is the truth;
  vertical line separates burn-in (20{,}000 iterations).
  \textit{Bottom:} gap $\overline{\log r}_t -
  \mathrm{ref\_logr}(\alpha_t)$; see text for interpretation.}
\label{fig:trace}
\label{fig:trace}
\end{figure}

\textbf{Smooth transition at the Cauchy boundary.}
Figure~\ref{fig:trace} displays the MCMC trace for one replication
at true $\alpha = 1.00$ ($n = 2{,}000$, $m = 30$).
The upper panel traces the sampled $\alpha$ values across 50{,}000
iterations; the chain crosses the Cauchy boundary $\alpha = 1$
a total of 190 times in the post-burn-in period, confirming that
the sampler explores both sides of this boundary freely and does
not become trapped in a sub-region of the parameter space.
The lower panel shows the gap
$\overline{\log r}_t - \mathrm{ref\_logr}(\alpha_t)$,
where $\overline{\log r}_t$ is the mean log-radial component
at iteration $t$ and $\mathrm{ref\_logr}(\alpha)
= \mathbb{E}_\alpha[\log r]$ is its expected value under the
truncated inverse-gamma distribution at the current $\alpha$.
Because $\mathrm{ref\_logr}(\alpha)$ is monotone increasing in
$\alpha$, the two panels move in opposition: a downward excursion
of $\alpha$ below the truth reduces the reference value by less
than it reduces $\overline{\log r}$, driving the gap negative,
while an upward excursion drives it positive; see also Theorem~\ref{thm:sign} in the online material.

\begin{figure}[t]
\centering
\includegraphics[width=0.75\textwidth]{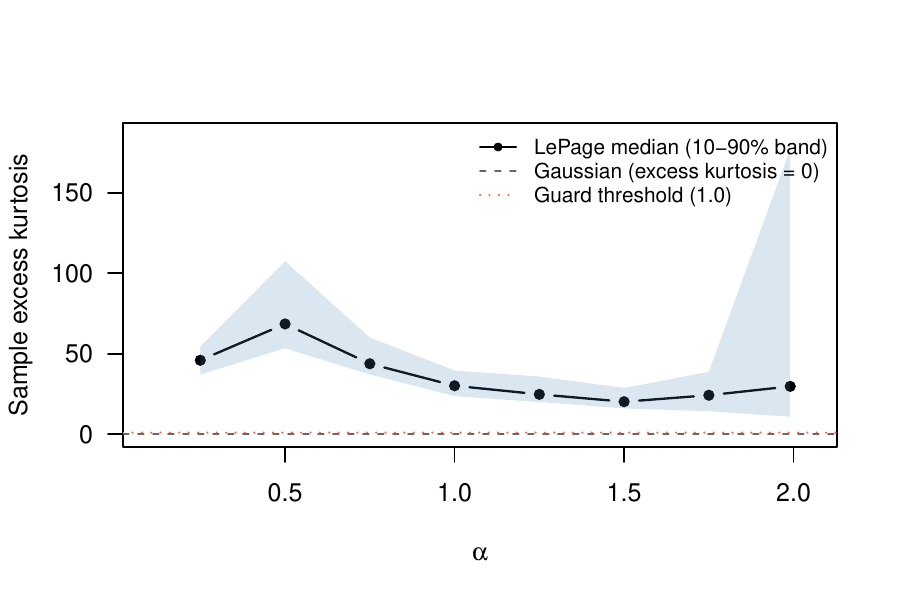}
\caption{Sample excess kurtosis of one marginal projection of
  LePage-generated bivariate stable data, as a function of the
  stability index $\alpha$ ($n=2000$, $m$ adaptive, 50 replications).
  Points show the median across replications; the shaded band spans
  the 10th to 90th percentiles.
  The dashed line marks the Gaussian benchmark (excess kurtosis~$=0$)
  and the dotted line marks the guard threshold of~$1.0$ used to
  detect near-Gaussianity.}
\label{fig:kurtosis_boundary}
\end{figure}

Figure~\ref{fig:kurtosis_boundary} plots the sample excess kurtosis of
one marginal projection of LePage-generated data against $\alpha$,
based on 50 replications at each grid point with $n=2000$ and adaptive
$m$.
Across the entire range $\alpha\in\{0.25,0.50,\ldots,1.99\}$ the
median excess kurtosis remains far above both the Gaussian benchmark of
zero and the guard threshold of $1.0$, the smallest median value being
approximately $20$ at $\alpha=1.50$.
This confirms that LePage samples at finite $m$ do not reproduce the
near-Gaussian tail behaviour of truly Gaussian data even at the
boundary $\alpha=1.99$.
The upturn in both the median and the band width at $\alpha=1.99$
reflects increasing variability of the LePage sum near the boundary.

\section{Discussion}
\label{sec:conclusion}

We have developed a full Bayesian inference framework for
$\alpha$-stable distributions with von Mises--Fisher spectral
measure in arbitrary dimension $d$, grounded in the LePage series
representation with latent radial, spectral, and Poisson
components. The likelihood, Gibbs sampler, and \texttt{logr\_gap}
diagnostic are derived for general $d$; the bivariate case $d=2$
with the circular von Mises measure serves as the simulation
testbed.

The sampler recovers all four parameters reliably across the full
range of $\alpha$, including the heavy-tailed extreme $\alpha=0.25$.
Continuity of the likelihood at the Cauchy boundary $\alpha=1$ is
established theoretically and confirmed empirically by free chain
crossing in both directions. Near the Gaussian boundary $\alpha=2$,
Fisher information for $\alpha$ collapses and ECF-based
initialisation can fail; the kurtosis-based guard provides a
data-adaptive remedy that eliminates catastrophic non-convergence
without affecting inference elsewhere.

Extension to $d>2$ requires no structural change: the radial and
Poisson draws are dimension-free, and the vMF spectral measure on
$\mathcal{S}^{d-1}$ replaces the circular von Mises distribution
directly. Only the projection integral requires numerical evaluation
for $d>2$.

We have shown that the von Mises -- Fisher spectral measure is supported by three independent lines of
argument. The regular variation characterisation
\eqref{eq:regular_variation} shows that $\kappa$ governs the
angular concentration of extreme observations independently
of~$\alpha$, giving the vMF family a geometrically transparent
parametrisation of tail anisotropy. Proposition~\ref{prop:anisotropy}
shows that the same $\kappa$ determines the eigenvalue ratio of the
Gaussian-boundary covariance, linking spectral concentration
monotonically to the shape of probability contours. Finally, the
hitting distribution of Brownian motion on $\mathcal{S}^{d-1}$
from an interior point is exactly vMF \citep{DGKendall1974, Gatto2013}, and
the associated pole-seeking diffusion \citep{DGKendall1974} has vMF as its stationary
distribution, providing a classical probabilistic foundation for
spectral concentration. Taken together, these connections place the
vMF spectral measure at the intersection of 
multivariate stable modelling and directional statistics.

\textbf{Disclaimer}

Views expressed are purely that of the author and do not reflect that of the organisation with which author is associated with.

\textbf{AI Policy Statement by the authors}

We have used Generative AI tools of Claude (Anthropic), version Claude Sonnet 4.6, accessed
via claude.ai. The AI assistant was used for LaTeX formatting, equation    typesetting, document    preparation,  grammar and style checking,
   R code debugging,   formatting of simulation scripts, and, 
  generation of initial drafts    of some remarks and captions. All scientific content, theoretical
results, simulation design,
interpretation of results, and
final decisions on all aspects of
the manuscript are solely the work
of the authors.  The AI tool did
not contribute to the intellectual
content of the paper.

\clearpage
\appendix

\begin{center}
  {\large \textbf{Supplementary Material - Theorems and proofs}}\\[6pt]
\end{center}


\section{Proof of Gaussian Boundary Theorem}

\begin{theorem}[Gaussian boundary, general $d$]
\label{thm:gaussian_boundary}
Let $X \sim \mathrm{Stable}(\alpha, \Gamma)$ in $\mathcal{R}^d$,
$\alpha \in (0,2)$, with finite spectral measure $\Gamma$ on
$S^{d-1}$ and total mass $\Gamma_\infty = \Gamma(S^{d-1}) < \infty$.
As $\alpha \to 2$,
\begin{equation}
  X \xrightarrow{d} N_d\!\left(0,\, \Sigma_\Gamma\right),
  \qquad
  \Sigma_\Gamma \;=\; 2\, \Gamma_\infty \,\int_{S^{d-1}} ss^\top\, \wG(ds).
  \label{eq:gaussian_boundary}
\end{equation}
The limiting covariance $\Sigma_\Gamma$ is determined by
the second moment of the normalised spectral measure
$\widetilde{\Gamma}$ and the total spectral
mass $\Gamma_\infty$; no further properties of $\Gamma$ are required.
\end{theorem}

\begin{proof}
\textbf{Step 1. Characteristic function.}
The characteristic function of an $\alpha$-stable
distribution in $\mathcal{R}^d$ with spectral measure $\Gamma$ is
\begin{equation}
  \log \varphi_X(t)
  \;=\; -\int_{S^{d-1}} |t^\top s|^\alpha
  \left[1 - i\,\mathrm{sgn}(t^\top s)
  \tan\!\left(\tfrac{\pi\alpha}{2}\right)\right]
  \Gamma(ds),
  \qquad t \in \mathcal{R}^d,
  \label{eq:cf_stable}
\end{equation}
(Samorodnitsky and Taqqu, 1994, Theorem~2.3.1), valid for all
$\alpha \in (0,2)$ and any finite spectral measure $\Gamma$.

\medskip
\textbf{Step 1b. Skewness term vanishes as $\alpha\to 2$.}
Since $|\mathrm{sgn}(t^\top s)| \leq 1$ for all $s \in S^{d-1}$
and $|t^\top s|^\alpha \leq \|t\|^\alpha$ on $S^{d-1}$,
\begin{equation}
  \left|
  \int_{S^{d-1}} |t^\top s|^\alpha\,
  \mathrm{sgn}(t^\top s)\,\Gamma(ds)
  \right|
  \;\leq\; \|t\|^\alpha\, \Gamma_\infty
  \;<\; \infty.
\end{equation}
Since $\tan(\pi\alpha/2) \to \tan(\pi) = 0$ as $\alpha \to 2$,
the imaginary part of $\log\varphi_X(t)$ satisfies
\begin{equation}
  \tan\!\left(\tfrac{\pi\alpha}{2}\right)
  \int_{S^{d-1}} |t^\top s|^\alpha\,
  \mathrm{sgn}(t^\top s)\,\Gamma(ds)
  \;\longrightarrow\; 0
  \qquad \text{as } \alpha \to 2.
\end{equation}
Consequently, the limit of $\log\varphi_X(t)$ as $\alpha\to 2$
is determined by the real part:
\begin{equation}
  \log \varphi_X(t)
  \;\sim\;
  -\int_{S^{d-1}} |t^\top s|^\alpha\, \Gamma(ds)
  \qquad \text{as } \alpha \to 2.
  \label{eq:cf_real}
\end{equation}

\medskip
\textbf{Step 2. Pointwise limit of the integrand.}
Fix $t \in \mathcal{R}^d$ and $s \in S^{d-1}$, and write
$u = |t^\top s| \geq 0$. Since $u^\alpha = \exp(\alpha \log u)$
is continuous in $\alpha$ for each fixed $u \geq 0$,
\begin{equation}
  |t^\top s|^\alpha \;\longrightarrow\; |t^\top s|^2
  \;=\; (t^\top s)^2
  \qquad \text{as } \alpha \to 2,
  \label{eq:pointwise}
\end{equation}
pointwise for every $s \in S^{d-1}$.

\medskip
\textbf{Step 3. Dominated convergence.}
For $\alpha \in [1,2]$, since $\|s\| = 1$ for all $s \in S^{d-1}$,
\begin{equation}
  |t^\top s|^\alpha
  \;\leq\; |t^\top s| + |t^\top s|^2
  \;\leq\; \|t\| + \|t\|^2.
  \label{eq:dombound}
\end{equation}
This bound is integrable with respect to $\Gamma$ since $\Gamma$
is finite:
\begin{equation}
  \int_{S^{d-1}}
    \bigl(|t^\top s| + |t^\top s|^2\bigr)\,\Gamma(ds)
  \;\leq\;
  \bigl(\|t\| + \|t\|^2\bigr)\, \Gamma_\infty
  \;< \infty.
\end{equation}
By the Dominated Convergence Theorem,
\begin{equation}
  \lim_{\alpha \to 2}
  \int_{S^{d-1}} |t^\top s|^\alpha\, \Gamma(ds)
  \;=\;
  \int_{S^{d-1}} (t^\top s)^2\, \Gamma(ds).
  \label{eq:dct}
\end{equation}

\medskip
\textbf{Step 4. Evaluate the limiting integral.}
Write $(t^\top s)^2 = t^\top s s^\top t$ and interchange
integration and the bilinear form:
\begin{equation}
  \int_{S^{d-1}} (t^\top s)^2\, \Gamma(ds)
  \;=\;
  t^\top \!\left(\int_{S^{d-1}} ss^\top\, \Gamma(ds)\right) t
  \;=\;
  t^\top M\, t,
  \label{eq:Mdef}
\end{equation}
where
\begin{equation}
  M \;=\; \int_{S^{d-1}} ss^\top\, \Gamma(ds)
\end{equation}
is a $d \times d$ positive semidefinite matrix (positive definite
whenever $\Gamma$ is not supported on a proper subspace of
$\mathcal{R}^d$).

\medskip
\textbf{Step 5. Identify the limiting characteristic function.}
Combining Steps~1b--4,
\begin{equation}
  \lim_{\alpha \to 2} \log \varphi_X(t)
  \;=\; -\,t^\top M\, t.
  \label{eq:cf_limit}
\end{equation}
For $Y \sim N_d(0, \Sigma)$, $\log \varphi_Y(t) =
-\tfrac{1}{2}\,t^\top \Sigma\, t$, so \eqref{eq:cf_limit}
is the log characteristic function of $N_d(0, 2M)$.

\medskip
\textbf{Step 6. Convergence in distribution.}
Since $\varphi_X(t) \to \exp(-t^\top M t)$ pointwise for all
$t \in \mathcal{R}^d$, and the limit is the characteristic function
of $N_d(0, 2M)$, which is continuous at $t = 0$, L\'{e}vy's
continuity theorem gives
\begin{equation}
  X \xrightarrow{d} N_d\!\left(0,\,\Sigma_\Gamma\right),
  \qquad
  \Sigma_\Gamma = 2M = 2 \, \int_{S^{d-1}} ss^\top\,\Gamma(ds).
\end{equation}
Writing $\Gamma = \Gamma_\infty\,\widetilde{\Gamma}$ with
$\widetilde{\Gamma}(S^{d-1}) = 1$, we have
$\Sigma_\Gamma = 2 \, \Gamma_\infty \, \int_{S^{d-1}} ss^\top\,
\widetilde{\Gamma}(ds)$, confirming that the limit is
determined by $\Gamma_\infty$ and the second moment of the
normalised spectral measure $\wG$.
\end{proof}

\begin{corollary}[von Mises--Fisher spectral measure]
\label{cor:vmf_covariance}

Let $\Gamma$ have density $\Gamma_\infty\, f_{\mathrm{vMF}}(\cdot\,;\mu,\kappa)$
with respect to the uniform surface measure on $S^{d-1}$,
$\mu \in S^{d-1}$ is the mean direction, and $\kappa \geq 0$ is
the concentration parameter. Then
\begin{equation}
  \Sigma_\Gamma
  \;=\; 2\Gamma_\infty
  \bigl[b(\kappa,d)\, \Id_d + c(\kappa,d)\,\mu\mu^\top\bigr],
\end{equation}
where
\begin{align}
  a(\kappa,d) &\;=\; \frac{1}{d}
    + \frac{d-1}{d}\,\frac{I_{d/2+1}(\kappa)}{I_{d/2-1}(\kappa)},
  \label{eq:akd} \\
  b(\kappa,d) &\;=\; \frac{1 - a(\kappa,d)}{d-1},
  \label{eq:bkd} \\
  c(\kappa,d) &\;=\; a(\kappa,d) - b(\kappa,d),
  \label{eq:ckd}
\end{align}
and $I_p(\cdot)$ denotes the modified Bessel function of the
first kind of order $p$. 
\end{corollary}

\begin{corollary}[Bivariate case, $d = 2$]
\label{cor:bivariate}
At $d = 2$, with $\mu = (\cos\eta, \sin\eta)^\top$ and
$R_2(\kappa) = I_2(\kappa)/I_0(\kappa)$,
Corollary~\ref{cor:vmf_covariance} reduces to
\begin{equation}
  \Sigma_\Gamma
  \;=\; \Gamma_\infty
  \!\left(\Id_2 + R_2(\kappa)\,A(\eta)\right),
  \qquad
  A(\eta) =
  \begin{pmatrix}
    \cos 2\eta & \sin 2\eta \\
    \sin 2\eta & -\cos 2\eta
  \end{pmatrix}.
\end{equation}
\end{corollary}

\begin{proof}
At $d = 2$, $I_{d/2+1}(\kappa)/I_{d/2-1}(\kappa) =
I_2(\kappa)/I_0(\kappa) = R_2(\kappa)$, so
$a(\kappa,2) = \tfrac{1}{2}(1 + R_2(\kappa))$,
$b(\kappa,2) = \tfrac{1}{2}(1 - R_2(\kappa))$,
$c(\kappa,2) = R_2(\kappa)$.
Substituting into Corollary~\ref{cor:vmf_covariance} and
using $2\mu\mu^\top = I_2 + A(\eta)$ (since
$2\cos^2\eta - 1 = \cos 2\eta$,
$2\sin^2\eta - 1 = -\cos 2\eta$,
$2\cos\eta\sin\eta = \sin 2\eta$) gives the result.
\end{proof}


\section{Monotonicity of the anisotropy ratio}

The matrix $\Sigma_\Gamma$ has
eigenvector $\mu$ with eigenvalue $2 \Gamma_\infty a(\kappa,d)$ and
all directions orthogonal to $\mu$ with eigenvalue
$2\, \Gamma_\infty b(\kappa,d)$; the anisotropy ratio
$\lambda_{\max}/\lambda_{\min} = a(\kappa,d)/b(\kappa,d)$
increases monotonically from $1$ (isotropic, $\kappa = 0$)
to $\infty$ (rank-$1$ degenerate, $\kappa \to \infty$).

\begin{proposition}
\label{prop:anisotropy_monotone}

Let
\[
a(\kappa,d)
=
\frac{1}{d}
+
\frac{d-1}{d}
\frac{I_{d/2+1}(\kappa)}
     {I_{d/2-1}(\kappa)}
\]
and
\[
b(\kappa,d)
=
\frac{1-a(\kappa,d)}{d-1},
\]
where $d\geq 2$ and $I_\nu$ denotes the modified Bessel function of the
first kind of order $\nu$. Define the anisotropy ratio by
\[
\rho(\kappa,d)
=
\frac{a(\kappa,d)}{b(\kappa,d)}
=
\frac{(d-1)a(\kappa,d)}
     {1-a(\kappa,d)}.
\]
Then $\rho(\kappa,d)$ is strictly increasing in $\kappa$ on
$(0,\infty)$. Moreover,
\[
\rho(0,d)=1
\]
and
\[
\lim_{\kappa\to\infty}\rho(\kappa,d)=\infty.
\]

\end{proposition}

\begin{proof}

Define
\[
R(\kappa,d)
=
\frac{I_{d/2+1}(\kappa)}
     {I_{d/2-1}(\kappa)}.
\]

\medskip
\noindent\textit{Step 1: $R(\kappa,d)$ is strictly increasing in
$\kappa$.}
Amos~\citep{Amos1974} shows that $\kappa\mapsto I_{\nu+2}(\kappa)/
I_\nu(\kappa)$ is strictly increasing on $(0,\infty)$ for all
$\nu\geq0$.
Setting $\nu=d/2-1$, which satisfies $\nu\geq0$ since $d\geq2$, gives
$R(\kappa,d)=I_{\nu+2}(\kappa)/I_\nu(\kappa)$, so $R(\kappa,d)$ is
strictly increasing on $(0,\infty)$.

\medskip
\noindent\textit{Step 2: $a(\kappa,d)$ is strictly increasing and
takes values in $(0,1)$ for $\kappa\in(0,\infty)$.}
Since
\[
a(\kappa,d)
=
\frac{1}{d}+\frac{d-1}{d}R(\kappa,d)
\]
and $(d-1)/d>0$, the function $a(\kappa,d)$ is strictly increasing in
$\kappa$.
For the bounds, note that $I_\nu(\kappa)>0$ for all $\kappa>0$ and
$\nu\geq0$, so $R(\kappa,d)>0$ and hence $a(\kappa,d)>1/d>0$.
For the upper bound, the modified Bessel function $I_\nu(\kappa)$ is
strictly decreasing in its order $\nu$ for fixed $\kappa>0$ and
$\nu\geq0$ \citep[\S10.37]{DLMF}, so $I_{d/2+1}(\kappa)<
I_{d/2-1}(\kappa)$ and therefore $R(\kappa,d)<1$.
Consequently,
\[
a(\kappa,d)
<
\frac{1}{d}+\frac{d-1}{d}
=1.
\]
Thus $a(\kappa,d)\in(0,1)$ for all finite $\kappa>0$, and
$\rho(\kappa,d)$ is well-defined.

\medskip
\noindent\textit{Step 3: $\rho(\kappa,d)$ is strictly increasing.}
Consider
\[
g(a)
=
\frac{(d-1)a}{1-a},
\qquad
0<a<1.
\]
Its derivative
\[
g'(a)
=
\frac{d-1}{(1-a)^2}>0
\]
shows that $g$ is strictly increasing on $(0,1)$.
Since $\rho(\kappa,d)=g(a(\kappa,d))$ and $a(\kappa,d)$ is strictly
increasing in $\kappa$, the composition $\rho(\kappa,d)$ is strictly
increasing on $(0,\infty)$.

\medskip
\noindent\textit{Step 4: Boundary values.}
For the lower boundary, the small-argument expansion
$I_r(\kappa)\sim(\kappa/2)^r/\Gamma(r+1)$ as $\kappa\to0^+$ gives
\[
\lim_{\kappa\to0^+}R(\kappa,d)
=
\lim_{\kappa\to0^+}
\frac{(\kappa/2)^{d/2+1}/\Gamma(d/2+2)}
     {(\kappa/2)^{d/2-1}/\Gamma(d/2)}
=
\lim_{\kappa\to0^+}
\frac{(\kappa/2)^2\,\Gamma(d/2)}{\Gamma(d/2+2)}
=0.
\]
Hence $a(0,d)=1/d$ and $b(0,d)=(1-1/d)/(d-1)=1/d$, so
$\rho(0,d)=1$.

For the upper boundary, the large-argument expansion
$I_r(\kappa)\sim e^\kappa/\sqrt{2\pi\kappa}$ as $\kappa\to\infty$
gives $\lim_{\kappa\to\infty}R(\kappa,d)=1$, so
\[
\lim_{\kappa\to\infty}a(\kappa,d)
=
\frac{1}{d}+\frac{d-1}{d}
=1,
\]
and therefore $\lim_{\kappa\to\infty}b(\kappa,d)=0$.
Consequently,
\[
\lim_{\kappa\to\infty}\rho(\kappa,d)
=
\lim_{\kappa\to\infty}
\frac{(d-1)\,a(\kappa,d)}{1-a(\kappa,d)}
=\infty.
\]
This completes the proof.
\end{proof}

\section{Derivation of the likelihood function in Section~\ref{sec:construction_likelihood} }
\label{sec:likelihood_derivation}

\begin{align}
  L(\alpha,\eta, \kap)
  &\propto f\!\left(\{X_i\},\{r_{ij}\},
    \{y_{ij}\},\{\bs_{ij}\}\mid
    \alpha,\eta, \kap\right)
  \nonumber\\
  &\propto
    f\!\left(\{r_{ij}\}\mid
    \alpha,\eta, \kap,\{X_i\}\right)
    \times
    f\!\left(\{y_{ij}\}\mid
    \alpha,\eta, \kap,\{X_i\},\{r_{ij}\}\right)
  \nonumber\\
  &\quad\times
    f\!\left(\{\bs_{ij}\}\mid
    \alpha,\eta, \kap,\{X_i\},\{r_{ij}\},
    \{y_{ij}\}\right)
    \times
    f\!\left(\{X_i\}\mid
    \alpha,\eta, \kap,\{r_{ij}\},\{y_{ij}\},
    \{\bs_{ij}\}\right)
  \nonumber\\[4pt]
  &=
    f\!\left(\{r_{ij}\}\mid\alpha\right)
    \times
    f\!\left(\{y_{ij}\}\mid
    \alpha,\{r_{ij}\}\right)
    \times
    f\!\left(\{\bs_{ij}\}\mid\eta, \kap\right)
  \nonumber\\
  &\quad\times
    f\!\left(\{X_i\}\mid
    \alpha,\eta, \kap,\{r_{ij}\},\{y_{ij}\},
    \{\bs_{ij}\}\right)
  \nonumber\\[4pt]
  &=
    f\!\left(\{r_{ij}\}\mid\alpha\right)
    \times
    f\!\left(\{y_{ij}\}\mid
    \alpha,\{r_{ij}\}\right)
    \times
    f\!\left(\{\bs_{ij}\}\mid\eta, \kap\right)
  \nonumber\\
  &\quad\times
    f\!\left(\{\beps_i\}\mid
    \alpha,\eta, \kap,\{r_{ij}\},\{y_{ij}\},
    \{\bs_{ij}\}\right)
  \nonumber\\[4pt]
  &=
    \left\{\prod_{i=1}^n\prod_{j=1}^m
    f(r_{ij}\mid\alpha)\right\}
    \times
    \left\{\prod_{i=1}^n\prod_{j=1}^m
    f(y_{ij}\mid\alpha,r_{ij})\right\}
  \nonumber\\
  &\quad\times
    \left\{\prod_{i=1}^n\prod_{j=1}^m
    f(\bs_{ij}\mid\eta, \kap)\right\}
    \times
    \left\{\prod_{i=1}^n
    f(X_i\mid\alpha,\eta, \kap,\{r_{ij}\},
    \{y_{ij}\},\{\bs_{ij}\})\right\}
  \nonumber\\[4pt]
  &=
    \left\{\prod_{i=1}^n\prod_{j=1}^m
    \frac{1}{\Gamma(\alpha)}
    \frac{e^{-1/r_{ij}}}
    {r_{ij}^{1+\alpha}}\right\}
    \times
    \left\{\prod_{i=1}^n\prod_{j=1}^m
    \frac{e^{-\lambda_{ij}}
    \lambda_{ij}^{y_{ij}}}
    {y_{ij}!}\right\}
  \nonumber\\
  &\quad\times
    \left\{\prod_{i=1}^n\prod_{j=1}^m
    f_Z(s_{ij}\mid\eta, \kap)\right\}
    \times
    \left\{\prod_{i=1}^n
    f(\beps_i\mid\alpha,\eta, \kap,
    \{r_{ij}\},\{y_{ij}\},
    \{\bs_{ij}\})\right\}.
\end{align}

\section{Smoothness of the Characteristic Function}
\label{sec:continuity_at_1}
\subsection{Continuity of the L\'evy-Khintchine integral}

Following Kuelbs, we express the characteristic function using the
Lévy--Khintchine integral. We show that this integral is smooth in the
stability index $\alpha\in(0,2)$, including at the Cauchy boundary
$\alpha=1$. Define
\begin{equation}
\label{eq:LK}
\K(v,\alpha)
=
\int_0^\infty
\left[
e^{irv}-1-\frac{irv}{1+r^2}
\right]
\frac{dr}{r^{1+\alpha}},
\qquad
v\in\mathcal{R},\quad \alpha\in(0,2).
\end{equation}

\begin{lemma}[Smoothness of the Lévy--Khintchine integral in $\alpha$]
\label{lem:LK_smooth}
For every $v\in\mathcal{R}$, the integral $\K(v,\alpha)$ in
\eqref{eq:LK} is finite for every $\alpha\in(0,2)$. Moreover, for each
fixed $v\in\mathcal{R}$, the map
$
\alpha\longmapsto \K(v,\alpha)
$
is $C^\infty$ on $(0,2)$. In particular, it is smooth at $\alpha=1$,
and no separate treatment of the Cauchy case is required.
\end{lemma}

\begin{proof}
Write
\[
\K(v,\alpha)=A(v,\alpha)+iB(v,\alpha),
\]
where
\begin{align}
A(v,\alpha)
&=
\int_0^\infty
\bigl[\cos(rv)-1\bigr]r^{-1-\alpha}\,dr,
\label{eq:A_lk}
\\
B(v,\alpha)
&=
\int_0^\infty
\left[
\sin(rv)-\frac{rv}{1+r^2}
\right]
r^{-1-\alpha}\,dr.
\label{eq:B_lk}
\end{align}

If $v=0$, then the integrand in \eqref{eq:LK} vanishes identically,
so that
\[
\K(0,\alpha)=0.
\]
Thus, assume for the moment that $v\neq0$.

As $r\to0^+$, Taylor expansion gives
\[
\cos(rv)-1=-\frac{r^2v^2}{2}+O(r^4)
\]
and
\[
\sin(rv)-\frac{rv}{1+r^2}=O(r^3).
\]
Consequently, near zero, the integrands in
\eqref{eq:A_lk} and \eqref{eq:B_lk} behave respectively as
\[
O(r^{1-\alpha})
\qquad\text{and}\qquad
O(r^{2-\alpha}).
\]
Both are integrable near zero because $\alpha<2$.

As $r\to\infty$,
\[
|\cos(rv)-1|\leq 2,
\]
and
\[
\left|
\sin(rv)-\frac{rv}{1+r^2}
\right|
\leq
1+\frac{|v|r}{1+r^2}
\leq 1+\frac{|v|}{r}.
\]
Therefore, both integrands are bounded at infinity by a constant
multiple of
\[
r^{-1-\alpha},
\]
which is integrable because $\alpha>0$. This proves that
$\K(v,\alpha)$ is finite for every
\[
(v,\alpha)\in\mathcal{R}\times(0,2).
\]

It remains to prove smoothness in $\alpha$. Fix
$v\in\mathcal{R}$ and let
\[
[\alpha_0,\alpha_1]\subset(0,2)
\]
be compact. For every integer $k\geq0$,
\[
\frac{\partial^k}{\partial\alpha^k}
\left[
\bigl(\cos(rv)-1\bigr)r^{-1-\alpha}
\right]
=
(-\log r)^k
\bigl(\cos(rv)-1\bigr)r^{-1-\alpha}.
\]

For $0<r<1$, the inequality
\[
|\cos(rv)-1|\leq \frac{v^2r^2}{2}
\]
gives
\[
\left|
(-\log r)^k
\bigl(\cos(rv)-1\bigr)r^{-1-\alpha}
\right|
\leq
\frac{v^2}{2}
|\log r|^k r^{1-\alpha_1}.
\]
Since $\alpha_1<2$, the function
\[
|\log r|^k r^{1-\alpha_1}
\]
is integrable on $(0,1)$.

For $r\geq1$, using $|\cos(rv)-1|\leq2$, we obtain
\[
\left|
(-\log r)^k
\bigl(\cos(rv)-1\bigr)r^{-1-\alpha}
\right|
\leq
2(\log r)^k r^{-1-\alpha_0}.
\]
This is integrable on $(1,\infty)$ because $\alpha_0>0$.

Thus, differentiation under the integral sign is justified, and
\[
\frac{\partial^k A}{\partial\alpha^k}(v,\alpha)
=
\int_0^\infty
(-\log r)^k
\bigl(\cos(rv)-1\bigr)r^{-1-\alpha}\,dr.
\]

The same argument applies to the imaginary part. Indeed,
\[
\frac{\partial^k B}{\partial\alpha^k}(v,\alpha)
=
\int_0^\infty
(-\log r)^k
\left[
\sin(rv)-\frac{rv}{1+r^2}
\right]
r^{-1-\alpha}\,dr.
\]
Near zero, the integrand is dominated by a constant multiple of
\[
|\log r|^k r^{2-\alpha_1},
\]
and near infinity by a constant multiple of
\[
(\log r)^k r^{-1-\alpha_0}.
\]
Both functions are integrable on their respective domains.

Therefore, $\K(v,\alpha)$ is $C^\infty$ as a function of $\alpha$ on
$(0,2)$. Since $1\in(0,2)$, this smoothness includes $\alpha=1$.
\end{proof}

\subsection{The Double Integral}

For $t \in \mathcal{R}^d$, write $t = \tau u$ where
$\tau = \|t\| \geq 0$ and $u = t/\|t\| \in S^{d-1}$
(for $t = 0$ the integral vanishes trivially).
For a $d$-dimensional stable model with vMF spectral density
$f_{\mathrm{vMF}}(s\mid\mu,\kappa)$ on $S^{d-1}$, the
relevant double integral is
\begin{equation}
\label{eq:double_integral}
\int_{S^{d-1}}
\int_0^\infty
\left[
e^{ir\langle t,s\rangle}
-1
-\frac{ir\langle t,s\rangle}{1+r^2}
\right]
\frac{dr}{r^{1+\alpha}}\,
f_{\mathrm{vMF}}(s\mid\mu,\kappa)\,ds.
\end{equation}

\begin{theorem}[Smoothness in the stability index, general $d$]
\label{thm:LK_smooth_d}
Fix $t \in \mathcal{R}^d$, $\mu \in S^{d-1}$, and
$\kappa \in [0,\infty)$. Then the double integral
\eqref{eq:double_integral} is $C^\infty$ as a function of
$\alpha \in (0,2)$, including at $\alpha = 1$.
\end{theorem}

\begin{proof}
For each $s \in S^{d-1}$, set $v(s) = \langle t, s\rangle$.
By the Cauchy--Schwarz inequality,
\[
  |v(s)| = |\langle t, s\rangle| \leq \|t\|\,\|s\| = \|t\|,
\]
so $v(s)$ is bounded uniformly in $s \in S^{d-1}$.
By Lemma~\ref{lem:LK_smooth}, the inner integral equals
$\mathcal{K}(v(s), \alpha)$ for each $s$, so
\eqref{eq:double_integral} equals
\[
  \int_{S^{d-1}}
  \mathcal{K}(v(s), \alpha)\,
  f_{\mathrm{vMF}}(s\mid\mu,\kappa)\,ds.
\]

If $t = 0$, then $v(s) = 0$ for every $s$ and
$\mathcal{K}(0,\alpha) = 0$, so the integral vanishes
identically and is trivially $C^\infty$.

Now suppose $t \neq 0$. Since $|v(s)| \leq \|t\|$
uniformly in $s$, the dominating functions in the proof of
Lemma~\ref{lem:LK_smooth} may be taken with $|v|$ replaced
by $\|t\|$, giving bounds that are uniform in $s$.
Specifically, for $\alpha \in [\alpha_0, \alpha_1]
\subset (0,2)$, the $k$th $\alpha$-derivative of the inner
integrand in \eqref{eq:double_integral} is dominated, near
$r = 0$, by a constant multiple of
\[
  |\log r|^k r^{1-\alpha_1}
  + |\log r|^k r^{2-\alpha_1},
\]
and, for $r \geq 1$, by a constant multiple of
\[
  (\log r)^k r^{-1-\alpha_0},
\]
both integrable over $(0,\infty)$ and independent of $s$.

Since $f_{\mathrm{vMF}}(\cdot\mid\mu,\kappa)$ is a
probability density on $S^{d-1}$,
\[
  \int_{S^{d-1}} f_{\mathrm{vMF}}(s\mid\mu,\kappa)\,ds = 1,
\]
the same bounds dominate the $s$-integrated expression.
By Fubini's theorem and differentiation under both integrals,
\eqref{eq:double_integral} is $C^\infty$ in $\alpha \in (0,2)$.
In particular, it is smooth at $\alpha = 1$.
\end{proof}

\section{Approximation with truncation}

\begin{proposition}[Truncation approximation]
\label{prop:truncation}
Let $X^{(m)}_{(\rmin,\rmax)}$ denote the LePage sum of
Theorem~\ref{thm:DLS_bivariate} with radial components
restricted to $(\rmin, \rmax)$, $0 < \rmin < \rmax
< \infty$, and with normalising constant
\[
  C_{\rmin}(\alpha)
  = \int_{\rmin}^{\rmax} r^{-\alpha-1}e^{-1/r}\,dr
\]
in place of $\Gamma(\alpha)$.
Then, for every fixed $m$ and every $t \in \mathbb{R}^d$,
\begin{equation}
  \psi^{(\rmin,\rmax)}_m(t)
  \;\longrightarrow\;
  \psi_m(t)
  \qquad
  \text{as } \rmin \to 0 \text{ and } \rmax\to\infty,
\end{equation}
where $\psi_m(t)$ is the characteristic function of
$X^{(m)}$ in Theorem~\ref{thm:DLS_bivariate}.
Consequently,
\begin{equation}
  \psi^{(\rmin,r_{\max})}_m(t)
  \;\longrightarrow\;
  \psi(t)
  \qquad
  \text{as } m\to\infty,\;
  \delta\to 0,\;
  r_{\max}\to\infty,
\end{equation}
where $\psi(t)$ is the characteristic function of
$\mathrm{Stable}(\alpha,\Gamma)$.
\end{proposition}

\begin{proof}
\textbf{Step 1. Convergence of the normalising constant.}
Since $r^{-\alpha-1}e^{-1/r} \geq 0$ and
\[
  \int_0^\infty r^{-\alpha-1}e^{-1/r}\,dr
  = \Gamma(\alpha) < \infty
  \quad \text{for } \alpha \in (0,2),
\]
the monotone convergence theorem gives
\begin{equation}\label{eq:Cdelta_limit}
  C_{\rmin}(\alpha)
  = \int_{\rmin}^{\rmax} r^{-\alpha-1}e^{-1/r}\,dr
  \;\longrightarrow\; \Gamma(\alpha)
  \qquad
  \text{as } \rmin\to 0,\; \rmax\to\infty.
\end{equation}

\medskip
\textbf{Step 2. Convergence of the truncated
L\'{e}vy--Khintchine integral.}
For $v \in \mathbb{R}$, define
\begin{equation}
  \mathcal{K}_{\delta,r_{\max}}(v,\alpha)
  = \int_\delta^{r_{\max}}
  \left[e^{irv} - 1 - \frac{irv}{1+r^2}\right]
  \frac{dr}{r^{1+\alpha}}.
\end{equation}
The integrand satisfies
\[
  \left|e^{irv}-1-\frac{irv}{1+r^2}\right|
  r^{-1-\alpha}
  \;\leq\;
  \begin{cases}
    \tfrac{v^2}{2}\,r^{1-\alpha} & r \leq 1, \\[4pt]
    C\,r^{-1-\alpha}             & r > 1,
  \end{cases}
\]
for a constant $C$ depending only on $v$ and $\alpha$.
Both bounds are integrable on $(0,\infty)$ for
$\alpha \in (0,2)$.  By the dominated convergence theorem,
\begin{equation}\label{eq:K_limit}
  \mathcal{K}_{{\rmin},r_{\max}}(v,\alpha)
  \;\longrightarrow\;
  \mathcal{K}(v,\alpha)
  \qquad
  \text{as } \rmin \to 0,\; r_{\max}\to\infty.
\end{equation}

\medskip
\textbf{Step 3. Convergence of the characteristic function.}
The log characteristic function of $X^{(m)}_{{\rmin},r_{\max}}$
takes the form
\[
  \log \psi^{({\rmin},r_{\max})}_m(t)
  = \Gamma_\infty
  \int_{S^{d-1}}
  \mathcal{K}_{{\rmin},r_{\max}}
  \bigl(\langle t, s\rangle,\alpha\bigr)\,
  \widetilde{\Gamma}(ds).
\]
By \eqref{eq:K_limit} and the fact that
$|\mathcal{K}_{\delta,r_{\max}}(\langle t,s\rangle,\alpha)|$
is bounded uniformly in $s \in S^{d-1}$ and in
$(\rmin, r_{\max})$ by an integrable function
(from Step~2 with $|v| \leq \|t\|$), a further application
of the dominated convergence theorem gives
\[
  \log \psi^{({\rmin},r_{\max})}_m(t)
  \;\longrightarrow\;
  \log \psi_m(t)
  \qquad
  \text{as } \rmin\to 0,\; r_{\max}\to\infty.
\]
The second statement follows by combining this with
Theorem~\ref{thm:DLS_bivariate}.
\end{proof}

\section{The logr\_gap Diagnostic}
\label{app:logr}

This appendix derives the theoretical
properties of the logr\_gap diagnostic
and proves the sign theorem used in
Section~\ref{sec:simulation}.

\subsection{The Truncated IG Distribution}

The radial draws $r_{ij}$ follow the
truncated inverted-gamma distribution
$\IG(\alpha,\rmin,\rmax)$ with density:
\begin{equation}\label{eq:IG_density_2}
  g(r\mid \alpha,\rmin,\rmax)
  = \frac{r^{-\alpha-1}e^{-1/r}}
    {\Cab(\alpha)},
  \quad r\in(\rmin,\rmax),
\end{equation}
where the normalising constant is:
\begin{equation}\label{eq:C_norm_2}
  \Cab(\alpha)
  = \int_{\rmin}^{\rmax}
    r^{-\alpha-1}e^{-1/r}\,dr.
\end{equation}
The reference log-r is:
\begin{equation}\label{eq:reflogr_def}
  \mathrm{ref\_logr}(\alpha)
  = E_{\IG(\alpha,\rmin,\rmax)}[\log r]
  = \frac{\int_{\rmin}^{\rmax}
    \log(r)\cdot r^{-\alpha-1}
    e^{-1/r}\,dr}{\Cab(\alpha)},
\end{equation}
and the logr\_gap diagnostic is:
\begin{equation}\label{eq:logr_gap}
  \mathrm{logr\_gap}
  = \overline{\log r}_{\rm chain}
  - \mathrm{ref\_logr}(\alpha_{\rm true}),
\end{equation}
where $\overline{\log r}_{\rm chain}$
is the post-burnin mean of $\log r_{ij}$.

\subsection{Exponential Family Structure}

\begin{proposition}[IG as exponential
  family]\label{prop:expfam}
  The density~\eqref{eq:IG_density_2} belongs
  to the one-parameter exponential family
  with natural parameter $\eta_{\rm ef}
  =-\alpha$, sufficient statistic
  $T(r)=\log r$, and log-partition function
  $A(\eta_{\rm ef})=\log\Cab(-\eta_{\rm ef})$.
\end{proposition}

\begin{corollary}[ref\_logr as log-partition
  derivative]\label{cor:mean_suff}
  By the standard exponential family identity:
  \begin{equation}\label{eq:reflogr_closed}
    \mathrm{ref\_logr}(\alpha)
    = -\frac{d}{d\alpha}\log\Cab(\alpha)
    = -\frac{\Cab'(\alpha)}{\Cab(\alpha)}.
  \end{equation}
\end{corollary}

\begin{proof}
  Under the substitution $t=1/r$:
  \begin{equation}
    \Cab(\alpha)
    = \int_{1/\rmax}^{1/\rmin}
      t^{\alpha-1}e^{-t}\,dt,
    \quad
    \Cab'(\alpha)
    = \int_{1/\rmax}^{1/\rmin}
      t^{\alpha-1}e^{-t}\log t\,dt,
  \end{equation}
  and $\int_{\rmin}^{\rmax}\log(r)
  r^{-\alpha-1}e^{-1/r}dr
  = -\Cab'(\alpha)$
  (since $\log r = -\log t$).
  Hence $\mathrm{ref\_logr}(\alpha)
  = -\Cab'(\alpha)/\Cab(\alpha)
  = -(d/d\alpha)\log\Cab(\alpha)$. 
\end{proof}

\subsection{Monotonicity and Fisher
  Information}

\begin{theorem}[Strict monotonicity]
\label{thm:monotone}
  $\mathrm{ref\_logr}(\alpha)$ is strictly
  decreasing in $\alpha$ on $(0,2)$:
  \begin{equation}\label{eq:deriv_reflogr}
    \frac{d}{d\alpha}
    \mathrm{ref\_logr}(\alpha)
    = -\Var_{\IG(\alpha,\rmin,\rmax)}
      [\log r] < 0.
  \end{equation}
\end{theorem}

\begin{proof}
  From the exponential family representation,
  the variance of the sufficient statistic
  $T(r)=\log r$ equals the second derivative
  of the log-partition function:
  \begin{equation}
    \Var_{\IG}[\log r]
    = \frac{d^2}{d\alpha^2}
      \log\Cab(\alpha) > 0.
  \end{equation}
  Differentiating~\eqref{eq:reflogr_closed}:
  \begin{equation}
    \frac{d}{d\alpha}
    \mathrm{ref\_logr}(\alpha)
    = -\frac{d^2}{d\alpha^2}
      \log\Cab(\alpha)
    = -\Var_{\IG}[\log r] < 0. 
  \end{equation}
\end{proof}

\begin{corollary}[Complete-data Fisher
  information]\label{cor:fisher}
  The Fisher information for $\alpha$
  from $nm$ iid draws from
  $\IG(\alpha,\rmin,\rmax)$ is:
  \begin{equation}\label{eq:fisher}
    \mathcal{I}(\alpha)
    = nm\cdot\Var_{\IG}[\log r]
    = nm\frac{d^2}{d\alpha^2}
      \log\Cab(\alpha) > 0,
  \end{equation}
  strictly positive for all
  $\alpha\in(0,2)$.  
\end{corollary}
\subsection{Proof of Proposition~\ref{prop:alpha_id}}

\begin{proof}
From the exponential family representation of
$\mathrm{IG}(\alpha,\rmin,\rmax)$
(Proposition~\ref{prop:expfam}),
the natural parameter is $\eta_{\rm ef}=-\alpha$ and the
sufficient statistic is $T(r)=\log r$.
By the standard exponential family identity, the variance of
the sufficient statistic equals the second derivative of the
log-partition function $A(\alpha)=\log C_{\rmin,\rmax}(\alpha)$:
\begin{equation}
  \Var_{\mathrm{IG}(\alpha,\rmin,\rmax)}[\log r]
  \;=\;
  \frac{d^2}{d\alpha^2}\log C_{\rmin,\rmax}(\alpha)
  \;=\;
  \frac{C_{\rmin,\rmax}''(\alpha)}{C_{\rmin,\rmax}(\alpha)}
  -\left(\frac{C_{\rmin,\rmax}'(\alpha)}
              {C_{\rmin,\rmax}(\alpha)}\right)^2
  \;>\;0,
\end{equation}
where the inequality holds since
$\mathrm{IG}(\alpha,\rmin,\rmax)$ is non-degenerate for
$0<\rmin<\rmax<\infty$.
The Fisher information for $\alpha$ from $nm$ independent
radial draws is therefore
\[
  \mathcal{F}_\alpha
  = nm\left[
      \frac{C_{\rmin,\rmax}''(\alpha)}{C_{\rmin,\rmax}(\alpha)}
      -\left(\frac{C_{\rmin,\rmax}'(\alpha)}
                  {C_{\rmin,\rmax}(\alpha)}\right)^2
    \right] > 0,
\]
which establishes \eqref{eq:FI_alpha}.
\end{proof}

\subsection{Connection to Score Equation}

The complete-data IG log-likelihood is:
\begin{equation}\label{eq:IG_lik}
  \ell_{\IG}(\alpha)
  = -nm\log\Cab(\alpha)
    - (1+\alpha)\sum_{i,j}\log r_{ij}
    - \sum_{i,j}r_{ij}^{-1}.
\end{equation}
Setting $\partial\ell_{\IG}/\partial\alpha
=0$ and using~\eqref{eq:reflogr_closed}:
\begin{equation}\label{eq:score_eq}
  \mathrm{ref\_logr}(\hat\alpha)
  = \overline{\log r}.
\end{equation}
The MCMC posterior mode satisfies this
exactly; the posterior mean satisfies it
approximately, with the approximation
improving as the posterior concentrates.

\subsection{The Sign Theorem}

\begin{theorem}[logr\_gap sign theorem]
\label{thm:sign}
  The posterior mean $\hat\alpha$ satisfies
  the approximate score
  equation~\eqref{eq:score_eq}.
  Therefore:
  \begin{equation}\label{eq:sign}
    \mathrm{sign}(\hat\alpha
      - \alpha_{\rm true})
    = -\mathrm{sign}(\mathrm{logr\_gap}),
  \end{equation}
  where $\mathrm{logr\_gap}=
  \overline{\log r}_{\rm chain}
  - \mathrm{ref\_logr}(\alpha_{\rm true})$.
\end{theorem}

\begin{proof}
  From~\eqref{eq:score_eq}:
  $\mathrm{ref\_logr}(\hat\alpha)
  \approx\mathrm{ref\_logr}(\alpha_{\rm true})
  +\mathrm{logr\_gap}$.
  If logr\_gap $>0$, then
  $\mathrm{ref\_logr}(\hat\alpha)
  >\mathrm{ref\_logr}(\alpha_{\rm true})$,
  and by strict decrease of
  $\mathrm{ref\_logr}$
  (Theorem~\ref{thm:monotone}),
  $\hat\alpha<\alpha_{\rm true}$
  (negative bias).  The case
  logr\_gap $<0$ gives positive bias
  by the same argument.  If logr\_gap
  $=0$ then $\hat\alpha=\alpha_{\rm true}$
  exactly. 
\end{proof}

\begin{remark}[Real-time diagnostic]
  Theorem~\ref{thm:sign} provides an
  immediate convergence check and bias
  predictor during MCMC:
  \begin{center}
  \renewcommand{\arraystretch}{1.3}
  \begin{tabular}{lll}
  \toprule
  logr\_gap & $\hat\alpha$ direction
    & Action \\
  \midrule
  $\approx 0$ & Unbiased & None \\
  $>0$ & Downward bias & Run longer \\
  $<0$ & Upward bias
    & Check misspecification \\
  \bottomrule
  \end{tabular}
  \end{center}
  The diagnostic is independent of
  $(\Psi,\eta,\kap,\sigma^2)$ and
  holds under correct specification and
  spectral misspecification alike.
\end{remark}

\subsection{Numerical Computation}

In practice, $\mathrm{ref\_logr}(\alpha)$
is computed via the change of variable
$t=1/r$:
\begin{equation}\label{eq:reflogr_numerical}
  \mathrm{ref\_logr}(\alpha)
  = \frac{D(\alpha,t_{\rm up})
    - D(\alpha,t_{\rm low})}
    {\gamma(\alpha,t_{\rm up})
    - \gamma(\alpha,t_{\rm low})},
\end{equation}
where $t_{\rm low}=1/\rmax$,
$t_{\rm up}=1/\rmin$,
$\gamma(\alpha,x)=\int_0^x
t^{\alpha-1}e^{-t}dt$ is the lower
incomplete Gamma function, and
$D(\alpha,x)=\partial\gamma(\alpha,x)
/\partial\alpha = \int_0^x t^{\alpha-1}
e^{-t}\log(t)dt$.  Both quantities are
available in standard numerical libraries.

\section{MCMC initialization}
\label{app:MCMC_initialisation}

\subsection{Overview}

The Gibbs sampler requires starting
values for four parameters:
$(\alpha, \kap, \eta, \Gamma_\infty)$.
All initialisations are derived from
the observed data $\{X_i\}_{i=1}^n$
using method-of-moments or
projection-based estimators, without
using the latent variables
$\{r_{ij}, y_{ij}, s_{ij}\}$ which
are not yet available at initialisation.

\subsection{Initialisation of $\alpha$}
\label{sec:alpha_init}

\subsubsection{Koutrouvelis ECF Estimator
  on Four Projections}

The stability index $\alpha$ is
initialised using the
\citet{Koutrouvelis1980} empirical
characteristic function (ECF)
estimator, applied independently
to four univariate projections of
the bivariate data $\{X_i\}
_{i=1}^n$, $X_i=(X_{i1},X_{i2})
^\top$:
\begin{equation}\label{eq:four_proj}
  Z_i^{(1)} = X_{i1},\quad
  Z_i^{(2)} = X_{i2},\quad
  Z_i^{(3)} = \frac{X_{i1}+X_{i2}}
    {\sqrt{2}},\quad
  Z_i^{(4)} = \frac{X_{i1}-X_{i2}}
    {\sqrt{2}}.
\end{equation}
These correspond to projections
onto unit vectors at angles
$0^\circ$, $90^\circ$, $45^\circ$, and $135^\circ$
respectively.  Each projected
series $\{Z_i^{(k)}\}$ is
univariate $\alpha$-stable with
the same stability index $\alpha$.

For each projection $k=1,\ldots,4$,
the Koutrouvelis estimator regresses
$\log(-\log|\hat\psi_n(t_j)|^2)$
against $\log|t_j|$ at a grid of
frequencies $\{t_j\}$:
\begin{equation}\label{eq:kout_reg}
  \log\bigl(-\log|\hat\psi_n(t_j)|^2
  \bigr)
  = \log(2\sigma^\alpha)
    + \alpha\log|t_j|
    + \epsilon_j,
\end{equation}
where $\hat\psi_n(t)=n^{-1}
\sum_{i=1}^n e^{itZ_i^{(k)}}$
and the slope gives
$\hat\alpha_{\rm Kout}^{(k)}$.

\begin{remark}[Choice of projection
  directions]
  The four directions $0^\circ$,
  $45^\circ$, $90^\circ$, $135^\circ$
  are chosen for three reasons.
  First, they form two orthogonal
  pairs covering the semicircle
  $[0^\circ,180^\circ)$ at equal
  spacing of $45^\circ$ --- projections
  beyond $180^\circ$ are redundant
  since $\langle-u,X\rangle$
  has the same stable index as
  $\langle u,X\rangle$.  Second,
  every possible mean direction
  $\eta\in[0^\circ,360^\circ)$ is
  within $22.5^\circ$ of one of the
  four directions, ensuring that at
  least one projection aligns
  reasonably well with the dominant
  tail direction regardless of
  $\eta$.  Third, the projections
  reduce to $X_{i1}$, $X_{i2}$,
  $(X_{i1}+X_{i2})/\sqrt{2}$, and
  $(X_{i1}-X_{i2})/\sqrt{2}$,
  which require no trigonometric
  computation.  The median over
  valid estimates provides
  robustness against any single
  poorly-aligned projection.
\end{remark}

\subsubsection{Robustness Steps}

Each individual estimate
$\hat\alpha_{\rm Kout}^{(k)}$
is validated before use:
\begin{enumerate}
  \item Extreme values of $\{Z_i^{(k)}\}$
    are trimmed to the
    $[0.1\%,99.9\%]$ quantile range
    before applying Kout.
  \item At least 50 observations
    and 20 distinct values are
    required; otherwise the
    estimate is discarded.
  \item The estimate is accepted
    only if finite and within
    $[\alpha_{\rm lower},
    \alpha_{\rm upper}]$; otherwise
    the fallback $\hat\alpha_0=0.70$
    is substituted.
\end{enumerate}

The final initialisation is the
median of all valid estimates:
\begin{equation}\label{eq:alpha_init}
  \hat\alpha_0 = \mathrm{median}
  \bigl\{\hat\alpha_{\rm Kout}^{(k)}:
  k\in\mathcal{V}\bigr\},
\end{equation}
where $\mathcal{V}\subseteq
\{1,2,3,4\}$ is the set of valid
projections ($|\mathcal{V}|\geq 2$
required; fallback $\hat\alpha_0
=0.70$ if $|\mathcal{V}|<2$).
The result is clipped to
$[\alpha_{\rm lower},
\alpha_{\rm upper}]$.

\begin{remark}[Robustness via
  four projections]
  Using four projections and taking
  the median provides robustness
  against outliers in any single
  projection direction.  The
  $45^\circ$ and $135^\circ$ projections
  are particularly useful when
  the data is strongly anisotropic
  in the axis directions.
\end{remark}

\begin{remark}[Fallback at the Gaussian boundary]
  Near $\alpha=2$, the Koutrouvelis estimator is known to be unstable
  \citep{Koutrouvelis1980} --- the regression~\eqref{eq:kout_reg}
  becomes nearly flat and the slope estimate unreliable.
  When all four projections fail or return out-of-range estimates,
  the fallback $\hat\alpha_0=0.70$ is used.
  To guard against this failure mode when data are known to be
  near-Gaussian, we additionally compute the sample excess kurtosis
  of the pooled projections: if it falls below $1.0$ --- inconsistent
  with $\alpha < 1.8$ at $n=2000$ --- the Koutrouvelis step is bypassed
  entirely and $\hat\alpha_0=1.95$ is used instead.
  This guard is activated only for the bivariate normal with covariance matrix induced by the von Mises spectral measure;  all other runs use the
  standard Koutrouvelis initialisation unchanged.
  Without this guard, one replication at true $\alpha=1.99$ (rep~202)
  received $\hat\alpha_0=0.70$, leading to non-convergence
  with posterior mean $\hat\alpha=1.530$ after 50{,}000 iterations.
  With the guard, rep~202 initialises at $\hat\alpha_0=1.95$ and
  converges to $\hat\alpha=1.993$.
\end{remark}

\begin{remark}[Software for Koutrouvelis initialisation]
  The iterative Koutrouvelis regression method
  \citep{Koutrouvelis1980} is implemented via the function
  \texttt{KoutParametersEstim()} in the \texttt{StableEstim}
  \textsf{R} package \citep{KharratBoshnakov2024}.
  The package also provides maximum likelihood, generalised method
  of moments, and Kogon--McCulloch estimators for univariate stable
  laws, but only the Koutrouvelis method is used here for
  initialisation.
\end{remark}
\subsection{Initialisation of $\kap$  and $\eta$}
\label{sec:kappa_eta_init}

\subsubsection{Unit-Vector Projections}

The spectral parameters $(\kap,\eta)$
are initialised from the directions
of the observed data.  For each
observation $X_i\in\R^2$ with
$\|X_i\|>0$, compute the unit
vector:
\begin{equation}\label{eq:unit_proj}
  u_i = \frac{X_i}{\|X_i\|}
  = \begin{pmatrix}
    \cos\hat\theta_i\\
    \sin\hat\theta_i
  \end{pmatrix},
  \quad
  \hat\theta_i = \mathrm{atan2}
  (X_{i2}, X_{i1})\in[0,2\pi).
\end{equation}
Observations with $\|X_i\|=0$ or
non-finite norms are excluded.

\subsubsection{Initial $\eta$: Circular  Mean Direction}

Compute the mean cosine and sine
of the projected directions:
\begin{equation}\label{eq:CS}
  \bar{C} = \frac{1}{n_0}
  \sum_{i\in\mathcal{U}}
  \cos\hat\theta_i,
  \qquad
  \bar{S} = \frac{1}{n_0}
  \sum_{i\in\mathcal{U}}
  \sin\hat\theta_i,
\end{equation}
where $\mathcal{U}$ is the set of
valid observations and
$n_0=|\mathcal{U}|$.
The mean resultant length and
initial mean direction are:
\begin{align}
  \hat{\Rbar}_0
  &= \sqrt{\bar{C}^2+\bar{S}^2},
  \label{eq:Rbar_init}\\
  \hat\eta_0
  &= \mathrm{atan2}(\bar{S},\bar{C})
  \;\mathrm{mod}\;2\pi.
  \label{eq:eta_init}
\end{align}

\subsubsection{Initial $\kap$:  Approximate Inversion of $\Rbar$}

The concentration $\kap$ is
initialised by inverting
$\Rbar(\kap)=I_1(\kap)/I_0(\kap)$
at $\hat{\Rbar}_0$.  Rather than
solving numerically, the following
piecewise approximation of
\citet{BestFisher1981} is used:
\begin{equation}\label{eq:kap_init}
  \hat\kap_0 = \begin{cases}
    \kap_{\rm lower}
    & \hat{\Rbar}_0 < 10^{-8},\\
    2\hat{\Rbar}_0
    + \hat{\Rbar}_0^3
    + \tfrac{5}{6}\hat{\Rbar}_0^5
    & \hat{\Rbar}_0 < 0.53,\\
    -0.4 + 1.39\hat{\Rbar}_0
    + \dfrac{0.43}{1-\hat{\Rbar}_0}
    & 0.53\leq\hat{\Rbar}_0<0.85,\\
    \dfrac{1}{\hat{\Rbar}_0^3
    -4\hat{\Rbar}_0^2
    +3\hat{\Rbar}_0}
    & \hat{\Rbar}_0\geq 0.85,
  \end{cases}
\end{equation}
clipped to
$[\kap_{\rm lower},\kap_{\rm upper}]
=[0.1, 50]$.

\begin{remark}[Approximation quality]
  The unit-vector projection
  $u_i=X_i/\|X_i\|$ extracts the
  direction of each observation
  but ignores its magnitude.  For
  heavy-tailed data ($\alpha$ small),
  the direction of $X_i$ is dominated
  by the largest radial component
  $r_{ij}$ and thus approximates
  the corresponding spectral angle
  $\theta_{ij}$.  For $\alpha$ near
  2, the data directions are more
  diffuse and the approximation
  is less accurate, but the Gibbs
  sampler corrects this during
  burn-in.
\end{remark}

\subsection{Initialisation of $\Gamma_\infty$}

\subsection{Eight-Direction Projection
  Estimator}

The total spectral mass $\Gamma_\infty$ is
initialised from the scale of
univariate projections across
$D=8$ equally spaced directions.
Let:
\begin{equation}\label{eq:dirs}
  \phi_k = \frac{(k-1)\pi}{D},
  \quad k=1,\ldots,D,
\end{equation}
with unit vectors $u_k=(\cos\phi_k,
\sin\phi_k)^\top$.  For each
direction $k$, project the
median-centred data:
\begin{equation}\label{eq:centred_proj}
  Z_i^{(k)} = \langle u_k,
  X_i - \tilde{X}\rangle,
  \quad i=1,\ldots,n,
\end{equation}
where $\tilde{X}=(\tilde{X}_1,
\tilde{X}_2)^\top$ is the
componentwise median.

\subsubsection{Per-Direction Scale
  Estimates}

For each direction $k$, apply the
Koutrouvelis estimator to
$\{Z_i^{(k)}\}$ to obtain
$\hat\alpha^{(k)}$ and scale
$\hat\gamma^{(k)}$ (the univariate
stable scale parameter).  The
stable scale satisfies:
\begin{equation}\label{eq:scale_rel}
  \hat\gamma^{(k)}
  = \bigl(\Gamma_\infty\,
    \II(u_k;\alpha,\eta,\kap)
    \bigr)^{1/\alpha},
\end{equation}
giving the per-direction $\Gamma_\infty$
candidate:
\begin{equation}\label{eq:zhat_k}
  \hat\Gamma_\infty^{(k)}
  = \frac{(\hat\gamma^{(k)})
    ^{\hat\alpha_{\rm use}}}
    {\hat\II(u_k;\hat\alpha_0,
    \hat\eta_0,\hat\kap_0)},
\end{equation}
where $\hat\alpha_0$ is the
estimate from
Section~\ref{sec:alpha_init}, and
$\hat\eta_0$, $\hat\kap_0$ are
from Section~\ref{sec:kappa_eta_init}.

\subsubsection{Numerical Evaluation of $\II$}

The scale integral for direction
$u_k=(\cos\phi_k,\sin\phi_k)^\top$
is evaluated numerically on a grid
of $G=4096$ equally spaced angles
$\theta_g=2\pi g/G$,
$g=0,\ldots,G-1$:
\begin{equation}\label{eq:I_numerical}
  \hat\II(u_k;\alpha,\eta,\kap)
  = \sum_{g=0}^{G-1}
    |\langle u_k,s(\theta_g)\rangle|
    ^\alpha
    f_{\vM}(\theta_g;\eta,\kap)
    \cdot\frac{2\pi}{G},
\end{equation}
where $s(\theta)=(\cos\theta,
\sin\theta)^\top$ and
$f_{\vM}(\theta;\eta,\kap)
=e^{\kap\cos(\theta-\eta)}
/(2\pi I_0(\kap))$.
Noting $\langle u_k,s(\theta_g)\rangle=\cos(\phi_k-\theta_g)$,
this simplifies to:
\begin{equation}\label{eq:I_numerical2}
  \hat\II(u_k;\alpha,\eta,\kap)
  = \frac{1}{G\,I_0(\kap)}
    \sum_{g=0}^{G-1}
    |\cos(\phi_k-\theta_g)|^\alpha
    \exp\!\bigl(\kap\cos(
    \theta_g-\eta)\bigr),
\end{equation}
a Riemann sum with $G$
points giving negligible
approximation error.

\subsubsection{Final Estimate}

Valid candidates are those with
$\hat\gamma^{(k)}>0$,
$\hat\II^{(k)}>0$, and
$\hat\Gamma_\infty^{(k)}\in(0,\infty)$.
The final initialisation is the
median over valid directions:
\begin{equation}\label{eq:zinfty_init}
  \hat\Gamma_{\infty ,0}
  = \mathrm{median}\bigl\{
    \hat\Gamma_\infty^{(k)}:
    k\in\mathcal{V}\bigr\},
\end{equation}
where $\mathcal{V}$ is the set of
valid directions, clipped to
$[z_{\rm lower}, z_{\rm upper}]
=[0.1, 50]$.

\begin{remark}[Robustness of the
  eight-direction estimator]
  Using $D=8$ directions and taking
  the median provides robustness
  against poor scale estimates in
  any single projection direction.
  If fewer valid directions are
  available, the median is taken
  over those that are valid.
  Errors in $(\hat\alpha_{\rm use},
  \hat\eta_0,\hat\kap_0)$ propagate
  to $\hat\Gamma_{\infty,0}$ through
  $\hat\II$, but are corrected
  during the 10{,}000-iteration
  burn-in.
\end{remark}

\bibliographystyle{plainnat}
\bibliography{references}
\end{document}